\documentclass[a4paper,scale=0.8,onecolumn,nobibnotes,nofootinbib]{revtex4}
\usepackage{enumitem}
\usepackage{amsmath,amssymb}
\usepackage{graphicx} 
\usepackage{epstopdf}
\usepackage{hyperref}
\usepackage{geometry}
\usepackage{subfigure}
\usepackage{makecell}
\usepackage{multirow}
\usepackage{float}
\usepackage{titlesec}
\usepackage{color}
\usepackage{pifont}
\usepackage{amssymb}
\usepackage{amsmath}
\usepackage{titlesec}
\usepackage{float}
\usepackage{caption}

\hypersetup{
    colorlinks=true,
    citecolor=blue,
    linkcolor=black,
    filecolor=magenta,
    urlcolor=cyan,}

\newcommand{\be}{\begin{equation}}
\newcommand{\ee}{\end{equation}}

\begin{document}

\title{ Thermodynamic topology of Kerr-Sen black holes  via R\'{e}nyi statistics}
\author{Meng-Yao Zhang \footnote{gs.myzhang21@gzu.edu.cn}$^{1}$, Hao Chen\footnote{ haochen1249@yeah.net}$^{2}$, Hassan Hassanabadi\footnote{h.hasanabadi@shahroodut.ac.ir}$^{3}$,  Zheng-Wen Long\footnote{zwlong@gzu.edu.cn (Corresponding author)}$^{4}$, and Hui Yang\footnote{huiyang@gzu.edu.cn (Corresponding author)}$^{1}$}
\affiliation{$^1$ School of Mathematics and Statistics, Guizhou University, Guiyang, 550025, China.\\
$^{2}$ School of Physics and Electronic Science, Zunyi Normal University, Zunyi 563006, China.\\
$^{3}$ Department of Physics, University of Hradec Kr\'{a}lov\'{e}, Rokitansk\'{e}ho 62, 500 03 Hradec Kr\'{a}lov\'{e}, Czechia.\\
$^{4}$ College of Physics, Guizhou University, Guiyang, 550025, China.
}

\date{\today}

\begin{abstract}
In the present study, we investigate the topological properties of black holes in terms of R\'{e}nyi statistics as an extension of the Gibbs-Boltzmann (GB) statistics, aiming to characterize the non-Boltzmannian thermodynamic topology of Kerr-Sen and dyonic Kerr-Sen black holes. Through this research, we discover that the topological number derived via R\'{e}nyi statistics differs from that obtained through GB statistics. Interestingly, although the nonextensive parameter $\lambda$ changes the topological number, the topological classification of the Kerr-Sen and dyonic Kerr-Sen black holes remains consistent under both GB and R\'{e}nyi statistics. In addition, the topological
numbers associated with these two types of black holes without cosmological constant using R\'{e}nyi entropy processes are the same as the AdS cases of them by considering the GB entropy, as further evidenced by such a study found here. This indicates the cosmological constant has some potential connections with the nonextensive parameter from the perspective of thermodynamic topology.
\end{abstract}
\maketitle

\section{Introduction}
The space-time of the Kerr-Sen black hole (BH) can be obtained from low energy limit of heterotic string theory. It exhibits different properties from those of general relativity, and its hidden conformal symmetries \cite{K1} and uniqueness \cite{K2} has been widely investigated. It is well known that Kerr-Sen black holes possess the rotation, charge and symmetry properties of asymptotically flat spacetime such as Kerr-Newman black holes. The physical properties of Kerr-sen solution exhibit certain similarities to those observed in the Einstein-Maxwell theory, yet it also demonstrates notable distinctions in various significant aspects. In addition to the U(1) vector field and metric tensor, this solution is characterized by two extra background fields: the 3-form H and the dilaton scalar field $\Phi$. The recent studies demonstrate that the weak cosmic censorship exists in Kerr-Sen black holes, even when taking into account the second-order perturbation inequality \cite{K3}; whereas, a (near-)extremal Kerr-Sen BH can be compromised by disregarding the self-force and radiative effects \cite{K4}. The Kerr-Newman spacetime and Kerr-Sen solution have been extensively compared in various aspects, including the shadows and evaporation of black holes \cite{K5,K6,K7}, gravitational capture regions of photons \cite{K8}, gyromagnetic ratios \cite{K9} and the superradiant instability of bound state with a rotating and electrically charged massive body \cite{K10,K11}. The dyonic Kerr-Sen black holes solution derived from the Einstein-Maxwell-Dilaton-Axion (EMDA) theory has been extensively investigated, including its AdS extension. This includes studies on the conformal invariance \cite{K12}, thermodynamic properties \cite{K13,K14}, black holes Shadows \cite{K15} and Chaos \cite{K16}.

On the other hand, base on the standard Gibbs-Boltzmann (GB) approach, a zero-charged BH in asymptotically flat space background in the phase with negative specific heat capacity, which means that it is unstable to reach thermal equilibrium \cite{R1}. In addition, Bekenstein's pioneering work \cite{R2} and the four laws of BH mechanics have been shown BH entropy is proportional to its area rather than its volume, which indicates that BH entropy should be taken as a nonextensive quantity in the BH backgrounds. The standard GB statistics may not be sufficient to calculate thermodynamic properties for extreme cases, such as BH systems with strong gravitational characteristics. In other words, the GB entropy formula exhibits limitations in the case of nonextensive long-range interaction BH system. The BH entropy $S_{BH}$ should encode the BH information with non-extensive nature, making a non-Boltzmannian statistics crucial. Several approaches, including holography \cite{R3,R4,R5}, string theory \cite{R6,R7,R8} and loop quantum gravity \cite{R9,R10} have been proposed as potential explanations for the non-extensive nature of BH entropy. Currently, the functional form of the BH entropy can potentially be expressed with respect to R\'{e}nyi statistics incorporating a non-extensive parameter, which reflects the non-extensivity exhibited by black holes. To address the stability issue, it is possible to achieve a stable BH by utilizing the R\'{e}nyi entropy instead of the conventional GB entropy \cite{R11,R12,R13,R14,R15,R16}. In addition, A possible relation between nonextensive parameter $\lambda$ and the cosmological constant $\Lambda$ first proposed in \cite{R14}, is strengthened in charged black holes \cite{R17, R18}, de Rham-Gabadadze-Tolley BH \cite{R19},  and Bardeen BH \cite{R20}, etc, further evidence is provided from the topological aspect \cite{R21, T26}.

Topology have emerged as an available mathematical tool for investigating physical systems, focusing on general characteristics of the system rather than specific structures. Defects play an important role in revealing certain properties of field configurations. Since Wei \cite{T1} identified BH solutions as thermodynamic defects through the establishment of topological numbers, the study of the thermodynamic topology of black holes has sparked research interest \cite{T2,T3,T4,T5,T6,Tx1,T7,T8,T9,T10,T11,T12,T13,T14,T15,T16,T17,T18,T19,T20,T21,T22,T23,T24,T25,Tx2,Tx3,Tx4,Tx5}. This includes exploring the topological structures of black holes in rotating spacetime and their corresponding AdS extension \cite{T7,T8,T9,T10,T11}, as well as examining complex BH context \cite{T18}. From a topological perspective, the positive and negative topological charges, namely the positive and negative winding numbers, can serve as indicators of local stability and instability characteristics of black holes. Furthermore, by summing up these topological charges, we can categorize BH solutions into three distinct classes to reflect the overall properties of the system. The results show that different black holes or even same ones may be are in the same class under different ensemble and parameter background; alternatively, they may belong to different topological class.

In this paper, we employ the nonextensive R\'{e}nyi statistics to study the thermodynamic topology of Kerr-Sen and dyonic Kerr-Sen black holes. We calculate the topological number under Boltzmannian statistics and non-Boltzmannian statistics, and we show that R\'{e}nyi parameter has a significant effect on the topological number. The outline of our paper is as follows: in section II, we begin with a review of thermodynamic topological method in Boltzmann statistics and introduce the generalized free energy represented by R\'{e}nyi entropy. Then we investigate the topological number of the Kerr-Sen BH in the absence of cosmological constant $\Lambda $  by utilizing the R\'{e}nyi statistics. In section III, we discuss the Kerr-Sen-AdS BH via conventional GB entropy. In section IV, the topological number of dyonic Kerr-Sen BH is calculated in the framework of R\'{e}nyi statistics. In section V, we extend the dyonic Kerr-Sen solution in section IV to the case of AdS with the GB statistics. Finally, our remarks and conclusions are given in section VI.
\section{Kerr-Sen BH via the R\'{e}nyi statistics}
Firstly, we investigate R\'{e}nyi thermodynamic topology of Kerr-Sen BH. Written in terms of the standard Boyer-Lindquist coordinates $(t,r,\theta, \phi)$, the Kerr-Sen metric can be expressed as \cite{k1,k2,k3}
\begin{equation}
\begin{aligned}
d s^2= & e^{-\Phi }\left \{ -\frac{\Delta _b}{\rho _b^2} (dt-a\sin ^2\theta d\phi)^2 +\frac{\rho _b^2}{\Delta _b}dr^2 \right \}  \\
& + e^{-\Phi }\left \{ \frac{\sin ^2\theta}{\rho _b^2}\left [ adt-(r^2+2br+a^2) d\phi\right ] ^2+ \rho _b^2d\theta ^2 \right \},  \\
\mathcal{B} =&\frac{2br}{\rho _b^2} a\sin ^2\theta dt\wedge d\varphi ,~A=-\frac{Qr}{\rho ^2} e^{-\Phi }(dt-a\sin ^2\theta d\phi),~e^{-\Phi }=\frac{\rho ^2}{\rho _b^2},  \\
\end{aligned}
\end{equation}
where
\begin{equation}
\begin{aligned}
\Delta_b=r^2+2(b-M) r+a^2, \rho_b^2=r^2+2 b r+a^2 \cos ^2 \theta,
\end{aligned}
\end{equation}
in which $b=Q^2/2M$ is the twist parameter, when $b=0$,  the above solution reduces to the Kerr geometry. $M$ is the BH mass, $J=Ma$ is the angular momentum, and $Q$ is the U(1) charge. Then we list the thermodynamic quantities such as mass, Hawking temperature and entropy of this BH as follows
\begin{equation}\label{M}
M=\frac{r_+^2+2br_++a^2}{2r_+},
\end{equation}
\begin{equation}
T_{BH}=\frac{r_+^2-a^2}{4\pi r_+(r_+^2+2br_++a^2)},
\end{equation}
\begin{equation}\label{S}
S_{BH}=\pi(r_+^2+2br_++a^2).
\end{equation}
In terms of GB statistics, the generalized off-shell Helmholtz free energy function can then be given by \cite{T1}
\begin{equation}
\mathcal{F }=M-\frac{S_{BH}}{\tau}.
\end{equation}
This free energy appears its off-shell characteristics except at $\tau=1/T_{BH}$. Based on the R\'{e}nyi statistics, we can rewrite the generalized off-shell free energy as follows
\begin{equation}\label{FR}
\mathcal{F_R } =M-\frac{S_{R}}{\tau_R}.
\end{equation}
The R\'{e}nyi entropy  $S_{R}$ can be expressed as $S_{R}=\frac{1}{\lambda}\mathrm {ln}(1+\lambda S_{BH})$ \cite{R6}. The acceptable range for the nonextensive parameter $\lambda$ is $-\infty <\lambda<1$ \cite{Ry}, otherwise the entropy function becomes ill-defined due to its convexity. In the current study on BH thermodynamics using R\'{e}nyi statistics, the entropy $S_R$ is always well defined for the parameter $\lambda$ within the range $0<\lambda<1$, with $\lambda$ exhibiting favorable thermodynamic properties in this interval, as demonstrated in \cite{R14}. When the R\'{e}nyi parameter $\lambda\to 0$, the generalized off-shell free energy returns to the GB statistics case. We establish a vector $\phi$ as \cite{T1}
\begin{equation}
\phi=(\frac{\partial \mathcal{F_R }}{\partial r_+}, -\mathrm { cot} \Theta \mathrm {csc} \Theta ),
\end{equation}
where the parameter $\Theta$ is an auxiliary item for convenient and intuitive topological analysis, it's introduction is for the purpose of axial limit \cite{tt}, and $\Theta$ obeys $0\le \Theta\le \pi$. Notably, when $\Theta=0,\pi$, the component $\phi^\Theta$ is divergent and the direction of the vector points is outward. Utilizing Duan's theory of $\phi$-mapping topological currents, we define a topological current as follows \cite{T27,T28,T29}
\begin{equation}
J^\mu=\frac{1}{2\pi } \epsilon ^{\mu \nu \sigma}\epsilon _{ab}\partial_{ \nu }n^a\partial_{ \sigma  }n^b,  \mu,\nu,\sigma =0,1,2,
\end{equation}
where $a,b=1,2$ and $\partial_{ \nu }=\frac{\partial}{\partial x^{\nu}}$, here $x^{\nu}=(\tau_R, r_+, \Theta)$. $n$ is a unit vector defined by $n=(n^1, n^2)$, where $n^1=\frac{\phi ^{r_+}}{\left \| \phi  \right \| }$, $n^2=\frac{\phi ^\Theta }{\left \| \phi  \right \| } $. The conservation of the topological current, i.e., $\partial_\mu J^\mu=0$, can be easily verified. Finally, the topological number in a parameter region $\sum $ is calculated as
\begin{equation}
\begin{aligned}
W  =\int_{\sum }^{} j^0d^2x=\sum_{i=1}^{N}  w_i,
\end{aligned}
\end{equation}
where $w_i$ is the winding number. In this section, we would like to calculate the topological number of the Kerr-Sen BH via R\'{e}nyi statistics. According to Eqs. \eqref{M}\eqref{S}\eqref{FR}. We obtain the generalized free energy as
\begin{equation}
\mathcal{F}_R =\frac{a^2+2b r_++r_+^2}{2r_+} -\frac{\mathrm {ln}  (1+\lambda \pi (a^2+2br_++r_+^2)) }{\lambda \tau _R},
\end{equation}
we construct a vector field $\phi$ whose components read
\begin{equation}\label{PR}
\phi ^{r_+}=\frac{r_+^2-a^2}{2r_+^2}  -\frac{2\pi(b+r_+)}{(1+\lambda \pi (a^2+2br_++r_+^2))\tau _R},
\end{equation}
\begin{equation}
\phi ^\theta =-\mathrm { cot} \Theta \mathrm {csc} \Theta.
\end{equation}
By considering the Eq. \eqref{PR} to be zero, a curve on the $r_+-\tau _R$ plane can be obtained. In the case of a Kerr-Sen BH, one can reach the following:
\begin{equation}
  \tau_R =\frac{4\pi r_+^2(b+r_+)}{(r_+^2-a^2)(1+\lambda \pi (a^2+2br_++r_+^2))}.
\end{equation}
We plot the curve between $r_+$ and $\tau_R$ with fixed $a/r_0=0.1$, $b/r_0=0.2$ and different values of $\lambda$ in FIG. 1.  Interestingly, annihilation/generation points occur at different values of $\lambda$. The critical nonextensive parameter $\lambda_c$ can be obtained with the condition
\begin{equation}\label{lc}
\frac{\partial\tau_R}{\partial r_+}=\frac{\partial^2\tau_R}{\partial r_+^2}=0.
\end{equation}
By solving this equation, we obtain $\lambda_c=0.57$. We previously mentioned that when the parameter $\lambda=0$, the generalized free energy returns to the GB statistical case, from FIG. 1(a), one generation point is located at $\tau_c/r_0=6.6523$. For $ \tau>\tau_c$, there are two BH branches, the solid red line is the stable BH branch, the dashed blue line is the unstable BH branch. In FIG. 2(a), we plot the unit vector field $n$ with $\tau/r_0=8$, the zero points can be found at $(r_+/r_0, \Theta)=(0.15,\pi/2)$, and $(r_+/r_0, \Theta)=(0.40,\pi/2)$, respectively (in this paper, the zero points of all vector fields $n$ are given from left to right). We have $w_1=1$, $w_2=-1$, the topological number of the Kerr-Sen BH in GB statistics is $W =0$. From FIG.1(b), when $\lambda<\lambda_c$, one generation and one annihilation points divide the BH into three branches, the stable small and large branches and the unstable intermediate branches, the generation and annihilation points are found at $\tau_a/r_0=6.3537$ and $\tau_b/r_0=11.3092$. Otherwise, i.e., $\lambda>\lambda_c$ (see FIG. 1(c)), there is only one stable BH branch. From FIG. 2(b) when choosing $\tau_a<\tau_R<\tau_b$ i.e., $\tau_R/r_0=9$, we find three zero points located at $(r_+/r_0, \Theta)=(0.14,\pi/2)$, $(r_+/r_0, \Theta)=(0.65,\pi/2)$ and $(r_+/r_0, \Theta)=(3.36,\pi/2)$. The corresponding winding numbers are $w_1=1$, $w_2=-1$ and $w_3=1$, hence again we have topological number $W=1$. For the case $\lambda>\lambda_c$, as shown in FIG. 2(c), only one zero point at $(r_+/r_0, \Theta)=(0.12,\pi/2)$, we have the same topological number $W=1$.
\begin{figure}[htbp]
	\centering
	\subfigure[$\lambda=0$]{
    \begin{minipage}[t]{0.31\linewidth}
		\centering
		\includegraphics[width=2.1in,height=1.8in]{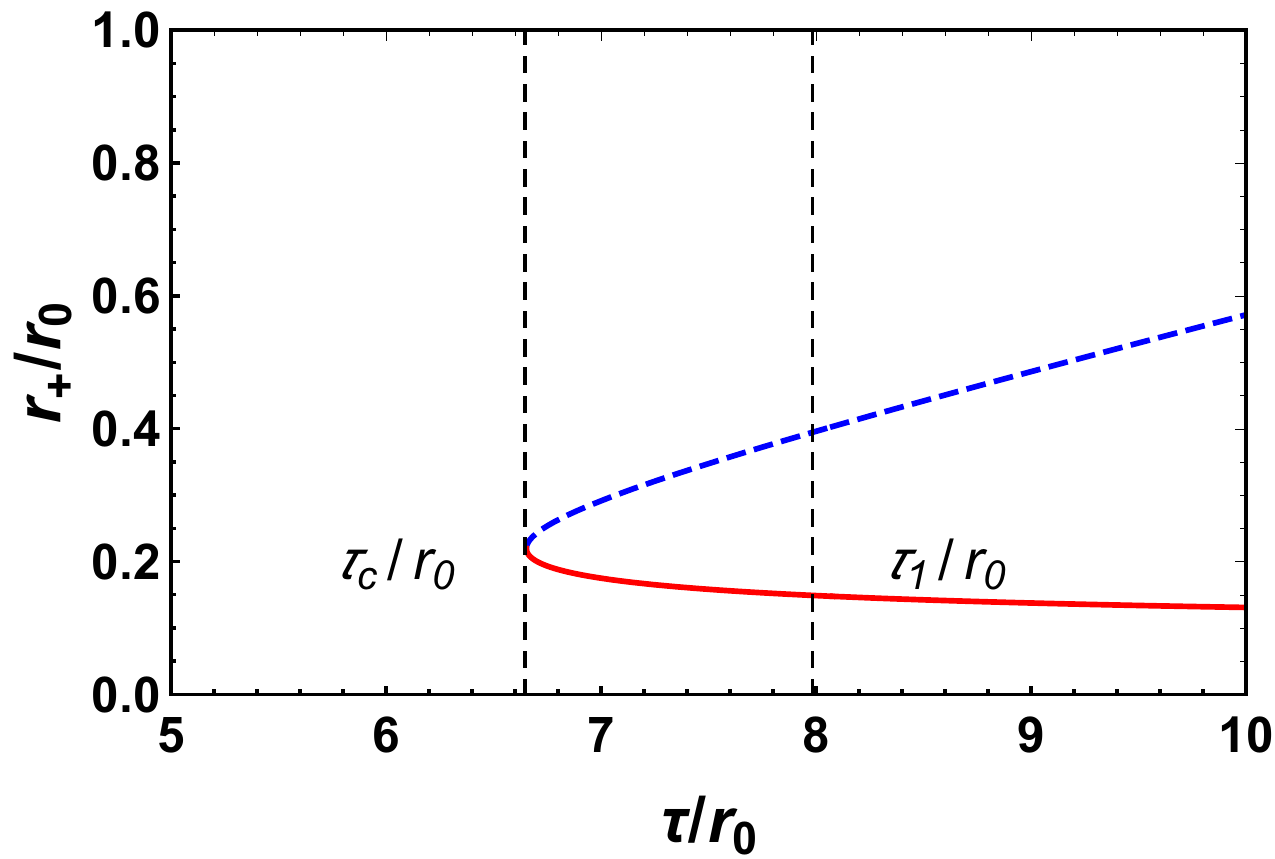}
		\end{minipage}%
            }%
    \subfigure[$\lambda<\lambda_c$]{
    \begin{minipage}[t]{0.31\linewidth}
		\centering
		\includegraphics[width=2.1in,height=1.8in]{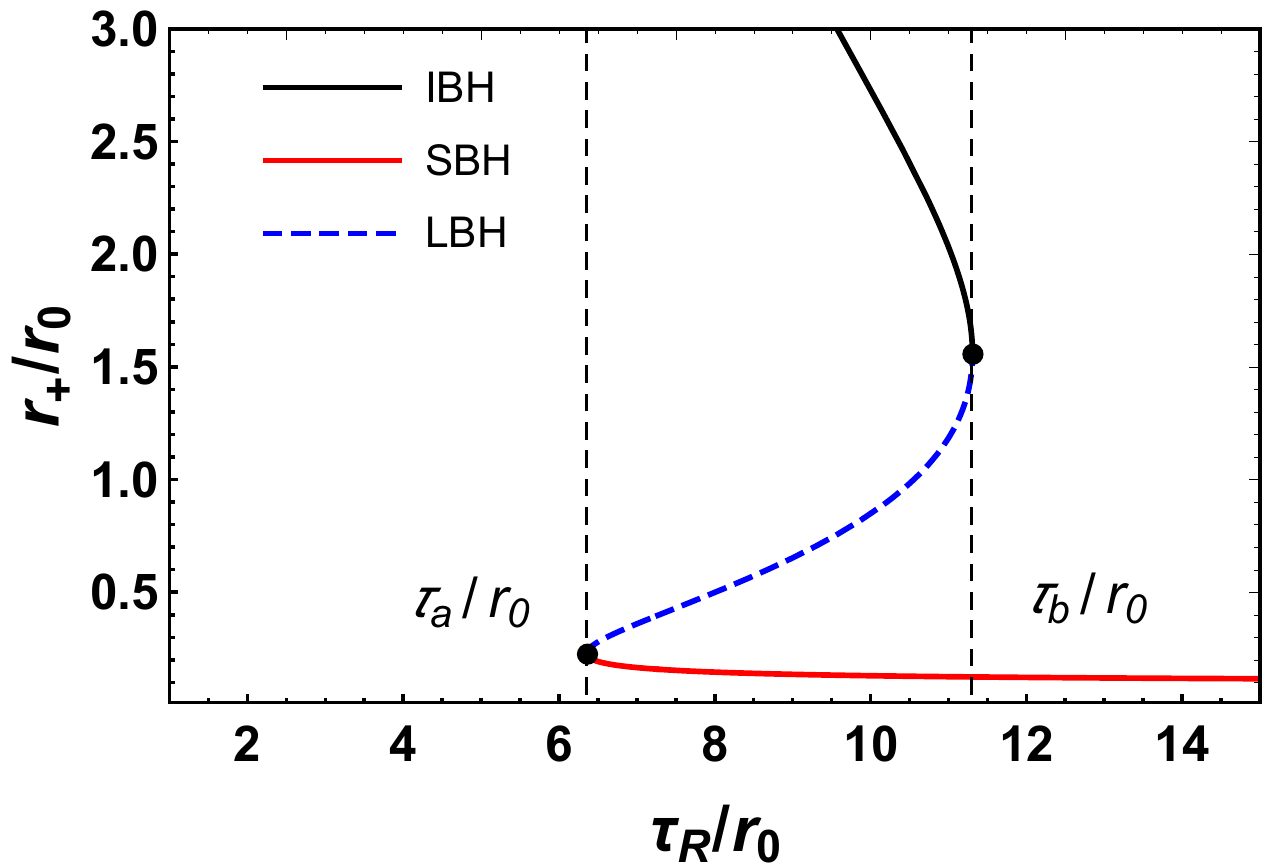}
		\end{minipage}%
            }%
      \subfigure[$\lambda>\lambda_c$]{
    \begin{minipage}[t]{0.31\linewidth}
		\centering
		\includegraphics[width=2.1in,height=1.8in]{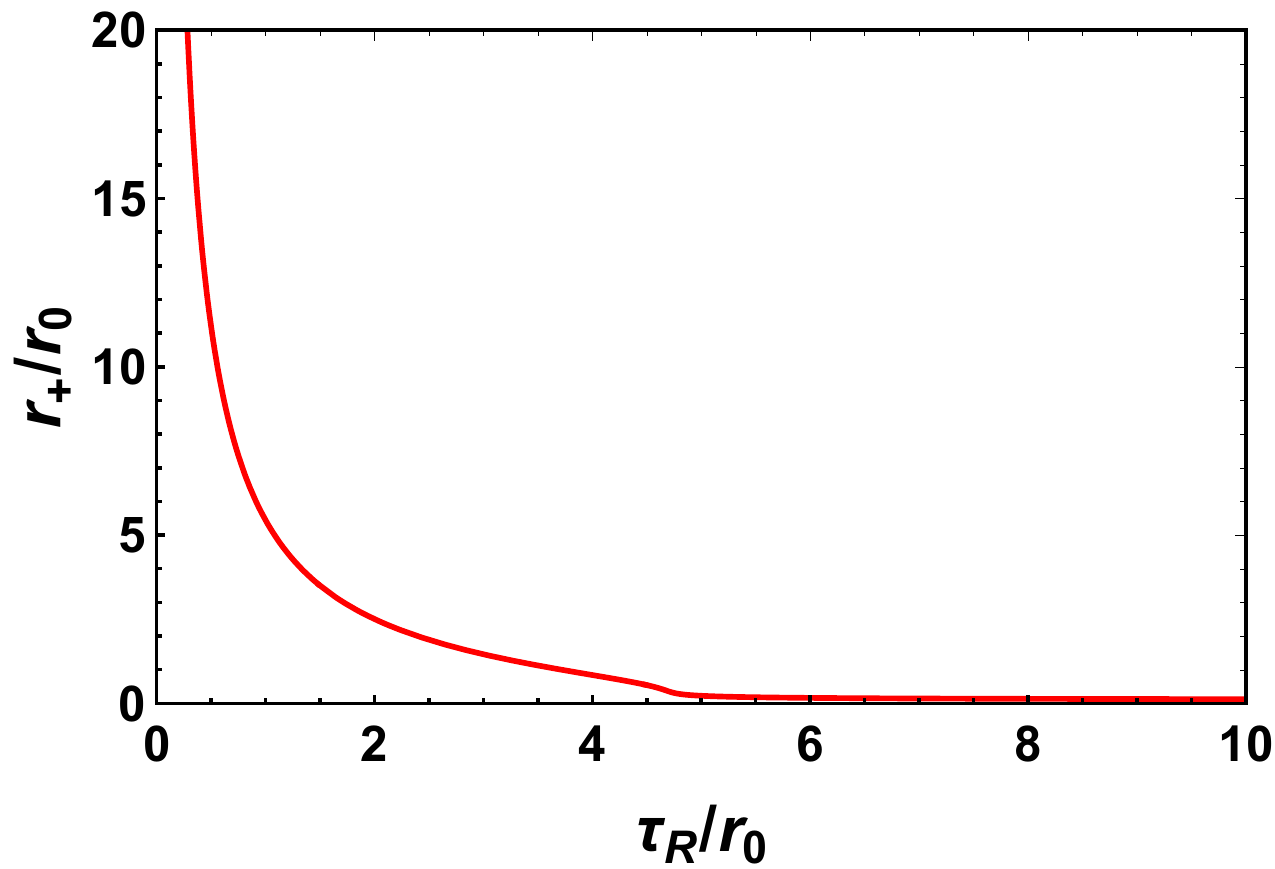}
		\end{minipage}%
            }%
     \centering
     \caption{The zero points of $\phi^{r_+}$ in the $\tau_R-r_+$ plane for Kerr-Sen BH via R\'{e}nyi statistics.  The left figure (a) is plotted with $\lambda=0$.  The middle figure (b) is plotted with $\lambda r_0^2=0.1$. The right figure (c) is plotted with $\lambda r_0^2=0.7$.}
\end{figure}

\begin{figure}[htbp]
	\centering
	\subfigure[$\lambda=0$]{
    \begin{minipage}[t]{0.31\linewidth}
		\centering
		\includegraphics[width=2.1in,height=1.8in]{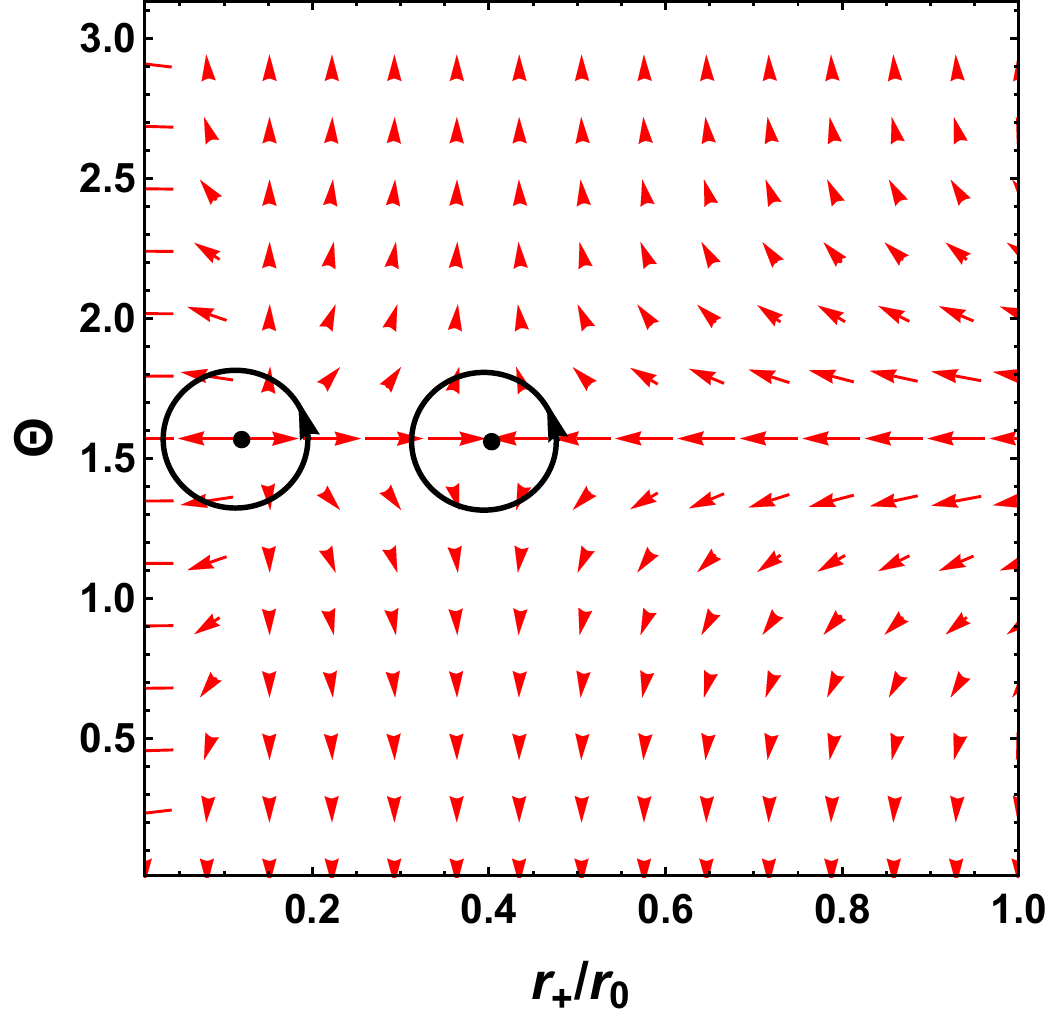}
		\end{minipage}%
            }%
    \subfigure[$\lambda<\lambda_c$]{
    \begin{minipage}[t]{0.31\linewidth}
		\centering
		\includegraphics[width=2.1in,height=1.8in]{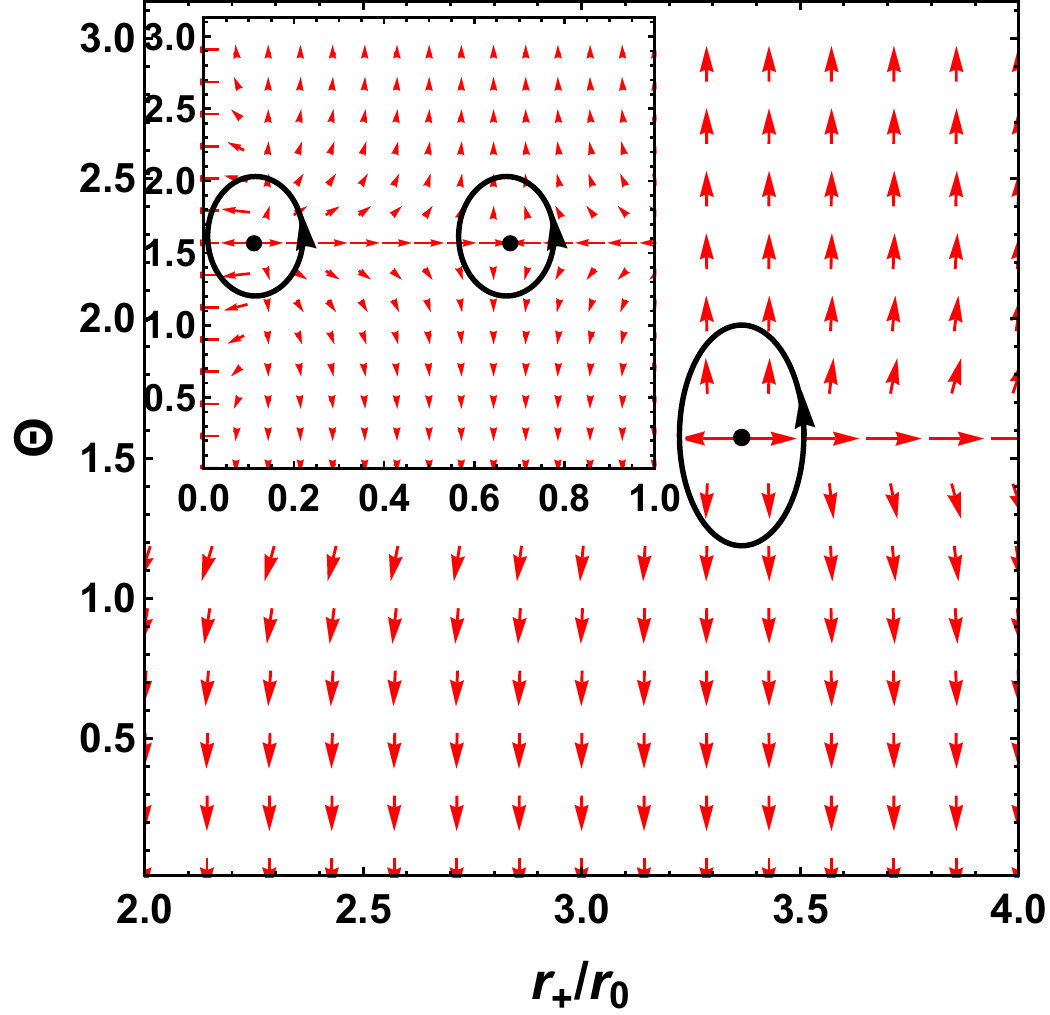}
		\end{minipage}%
            }%
      \subfigure[$\lambda>\lambda_c$]{
    \begin{minipage}[t]{0.31\linewidth}
		\centering
		\includegraphics[width=2.1in,height=1.8in]{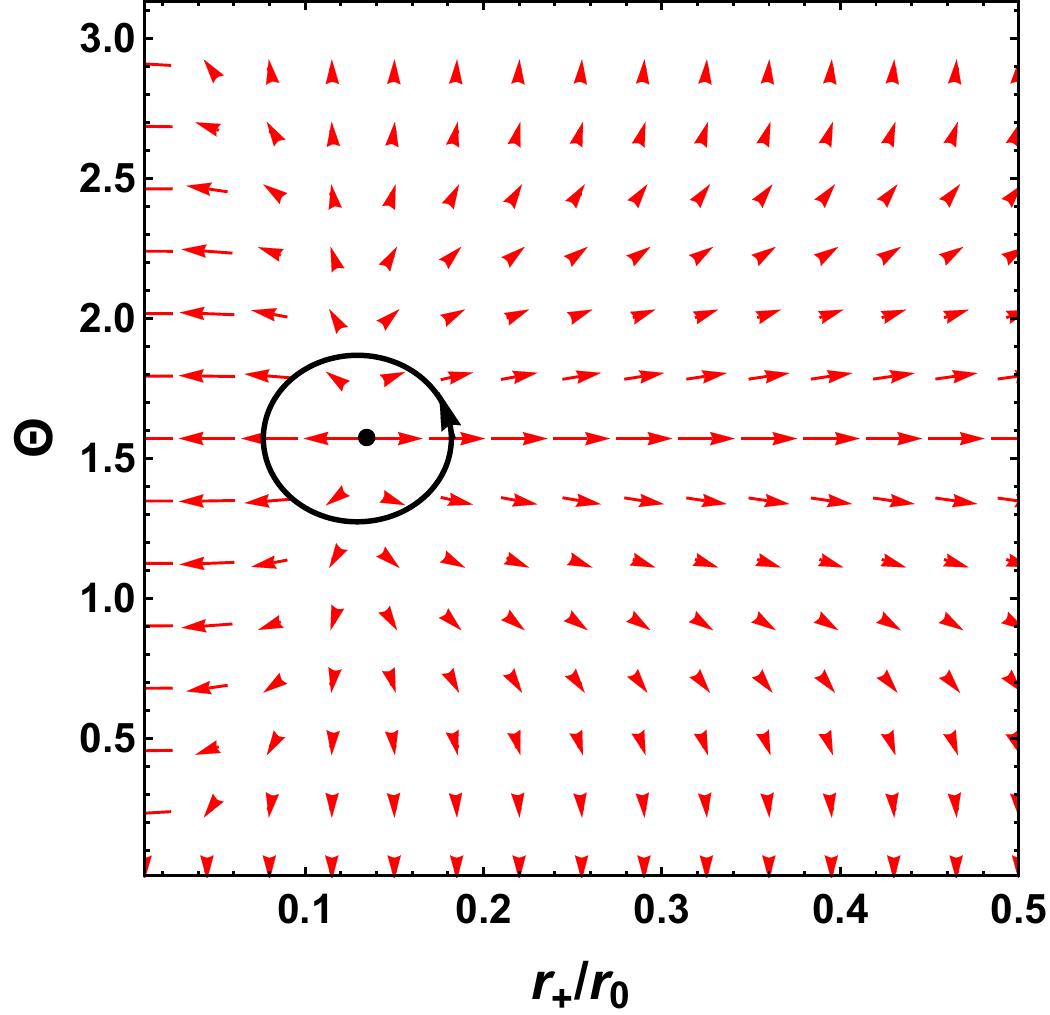}
		\end{minipage}%
            }%
     \centering
     \caption{The unit vector field $n$ on the $r_+-\Theta$ plane for Kerr-Sen BH via R\'{e}nyi statistics.  The left figure (a) is plotted with $\lambda=0$.  The middle figure (b) is plotted with $\lambda r_0^2=0.1$. The right figure (c) is plotted with $\lambda r_0^2=0.7$.}
\end{figure}
\section{Kerr-Sen-AdS BH via GB statistics}
In Boyer-Lindquist coordinates, the Kerr-Sen-AdS BH solution \cite{ka1} is expressed as
\begin{equation}
\begin{aligned}
d s^2= & -\frac{\Delta_r}{\Omega}\left(d t-\frac{a \sin ^2 \theta}{k} d \phi\right)^2+\frac{\Omega}{\Delta_r} d r^2+\frac{\Omega}{\Delta_\theta} d \theta^2 \\
& +\frac{\Delta_\theta \sin ^2 \theta}{\Omega}\left(a d t-\frac{\left(r^2+2 b r+a^2\right)}{k} d \phi\right)^2,
\end{aligned}
\end{equation}
where
\begin{equation}
\begin{aligned}
& \Delta_r=\left(r^2+2 b r+a^2\right)\left(1+\frac{r^2+2 b r}{l^2}\right)-2 m r, \quad k=1-\frac{a^2}{l^2}, \\
& \Omega=r^2+2 b r+a^2 \cos ^2 \theta, \quad \Delta_\theta=1-\frac{a^2}{l^2} \cos ^2 \theta .
\end{aligned}
\end{equation}
Through the horizon constraint can be derived by solving the equation $\Delta_r=0$. Then the mass and other thermodynamic variables can be expressed by the radius of the event horizon
\begin{equation}
M=\frac{m}{k^{2}},
\end{equation}
\begin{equation}
T_{BH}=\frac{a^2(r_+^2-l^2)+r_+^2((2b+r_+)(2b+3r_+)+l^2)}{4\pi r_+(a^2+2br_++r_+^2) l^2},
\end{equation}
\begin{equation}
S_{BH}=\frac{\pi(a^2+2br_++r_+^2)}{\quad k},
\end{equation}
the generalized free energy in GB statistics takes the form
\begin{equation}
\mathcal{F} =\frac{3(a^2+r_+(2b+r_+))(16\pi^2Pa^2r_++3\tau +2\pi r_+(8bP\tau +4Pr_+\tau -3))}{2(3-8\pi Pa^2)^2r_+\tau },
\end{equation}
\begin{figure}[htbp]
	\centering
    \subfigure[]{
    \begin{minipage}[t]{0.4\linewidth}
		\centering
		\includegraphics[width=2.5in,height=2.0in]{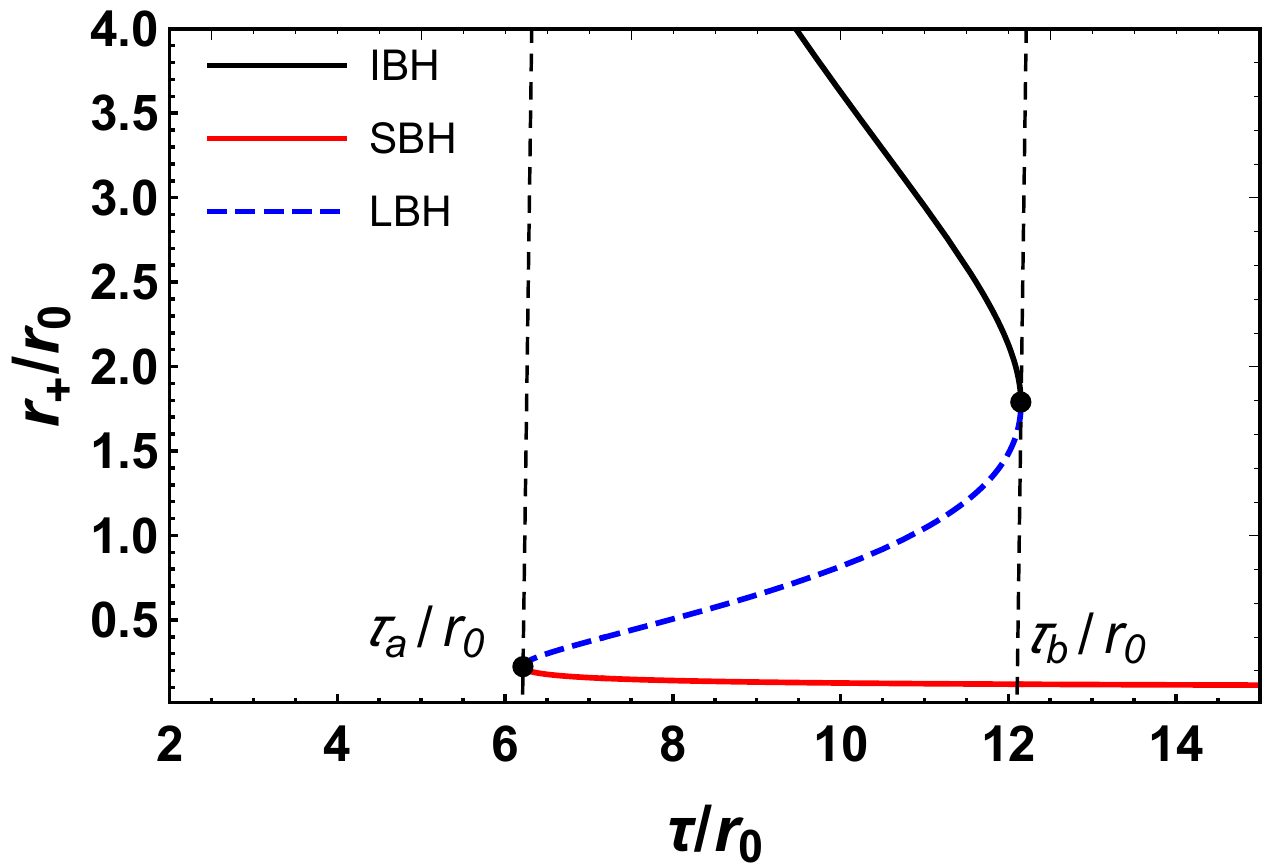}
		\end{minipage}%
            }%
      \subfigure[]{
    \begin{minipage}[t]{0.4\linewidth}
		\centering
		\includegraphics[width=2.5in,height=2.0in]{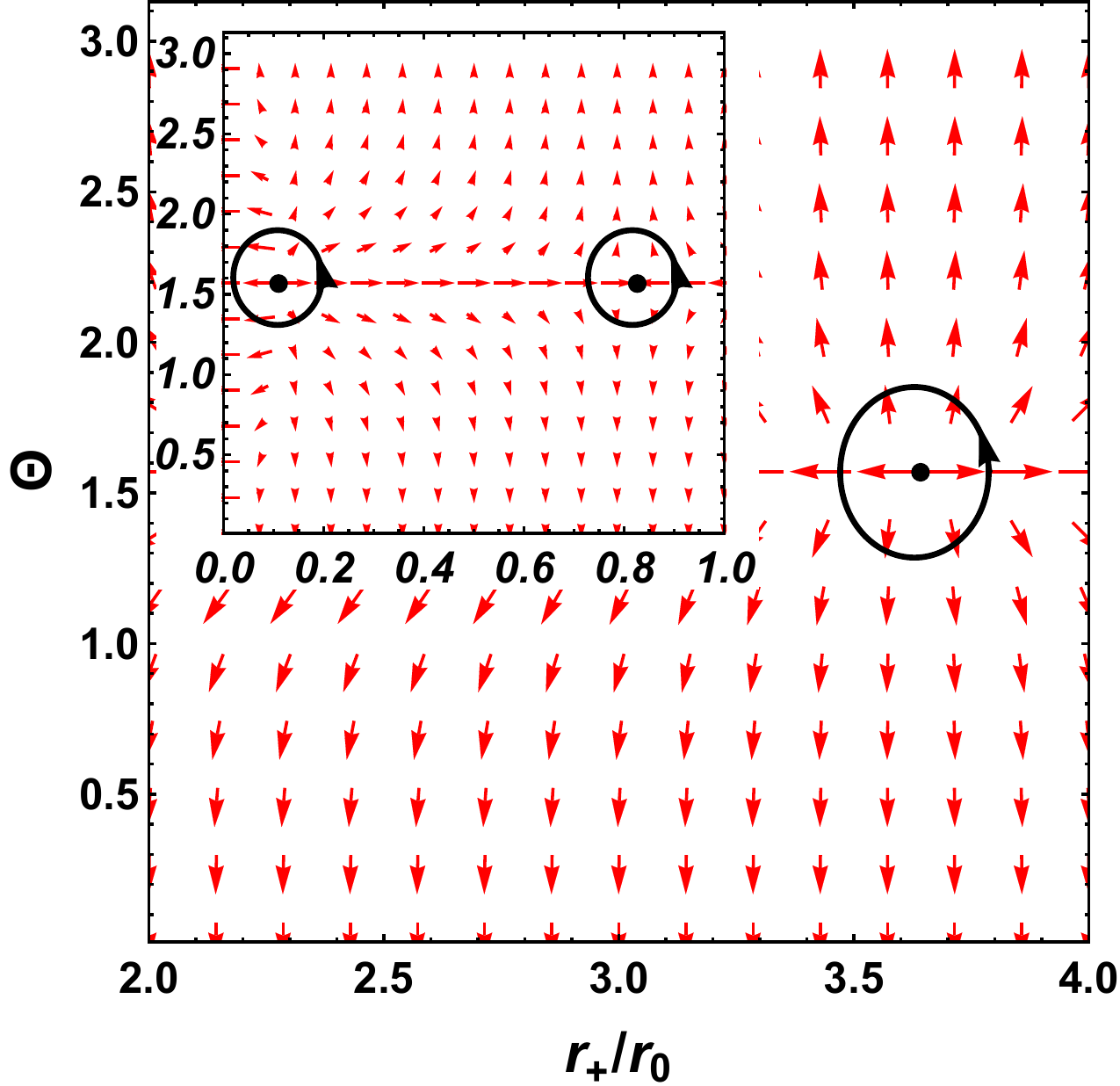}
		\end{minipage}%
            }%
     \centering
     \caption{Topological properties for Kerr-Sen-AdS BH via GB statistics. The left figure (a) reprensents the zero points of $\phi^{r_+}$ in $\tau-r_+$ plane. The right figure (b) reprensents the unit vector field $n$ on the $r_+-\Theta$ plane with $\tau/r_0=10$. }
\end{figure}
where $P=\frac{3}{8\pi l^2}$ is the thermodynamic pressure. According to the definition of the vector $\phi$, its components are
\begin{equation}
\phi^{r_+}=\frac{12\pi(8\pi Pa^2-3)(b+r_+)r_+^2+3(a^2(8\pi Pr_+^2-3)+r_+^2(3+8\pi P(2b+r_+)(2b+3r_+)))\tau }{2(3-8\pi Pa^2)^2r_+^2\tau },
\end{equation}
\begin{equation}
\phi ^\theta =-\mathrm { cot} \Theta \mathrm {csc} \Theta,
\end{equation}
by solving $\phi^{r_+}=0$ , we can obtain
\begin{equation}\label{t2}
\tau =-\frac{4\pi (8\pi Pa^2-3)(b+r_+)r_+^2}{a^2(8\pi Pr_+^2-3)+r_+^2(3+8\pi P(2b+r_+)(2b+3r_+))}.
\end{equation}
The graph of $r_+$ vs $\tau$ is plotted in FIG. 3(a) for $a/r_0=0.1$, $b/r_0=0.2$ and $Pr_0^2=0.01$ (below critical pressure). Interestingly, we find the curves of equation \eqref{t2} look similar to the case of Kerr-Sen BH via R\'{e}nyi statistics with $\lambda<\lambda_c$. The generation and annihilation points are obtained at $\tau_a/r_0=6.2078$ and $\tau_b/r_0=12.1509$. FIG. 3(b) represents the unit vector field $n$ with $\tau/r_0=10$. The zero points are at $(r_+/r_0, \Theta)=(0.13,\pi/2)$, $(r_+/r_0, \Theta)=(0.82,\pi/2)$ and $(r_+/r_0, \Theta)=(3.63,\pi/2)$, respectively. Thus the topological number of Kerr-Sen-AdS BH via GB statistics is 1, which is same as the Kerr-Sen BH from R\'{e}nyi statistics.
\section{Dyonic Kerr-Sen BH via the R\'{e}nyi statistics}
The EMDA theory is considered to find dyonic Kerr-Sen BH, the corresponding BH spacetime metric can be written as follows: \cite{K14}
\begin{equation}
\begin{aligned}
d \hat{s}^2  =-\frac{\hat{\Delta}(r)}{\hat{\Sigma}} \hat{X}^2+\frac{\hat{\Sigma}}{\hat{\Delta}(r)} d r^2+\hat{\Sigma} d \theta^2+\frac{\sin ^2 \theta}{\hat{\Sigma}} \hat{Y}^2,
\end{aligned}
\end{equation}
where
$$
\begin{aligned}
& \hat{X}=d t-a \sin ^2 \theta d \hat{\varphi}, \quad \hat{Y}=a d t-\left(r^2-2 d r-k^2+a^2\right) d \hat{\varphi}, \\
& \hat{\Delta}(r)=r^2-2 d r-2 m(r-d)-k^2+a^2+p^2+q^2, \\
& \hat{\Sigma}=r^2-2 d r-k^2+a^2 \cos ^2 \theta,
\end{aligned}
$$
where parameters $m$, $q$, $a$, $k$, $d$, $p$ represent mass, electric charge, spin, axion charge, dilaton charge, and magnetic (dyonic) charge of the BH, respectively. Here, $d=\frac{p^2-q^2}{2m}$, and $k=\frac{pq}{m}$, so when $p=q$, the dilaton charge $d$  will vanish, and when $p=0$, one have a vanishing axion charge $k$ ($k=0$), in this case the Kerr-Sen BH can be reduced. The Kerr BH solution can be obtained by considering $p=q=0$. In the following calculation, we set $p\ne q\ne 0$. The expression for $m$ can be derived by $\Delta=0$. The other quantities are given by
\begin{equation}
M=m,
\end{equation}
\begin{equation}
\begin{aligned}
T_{BH}  =\frac{r_{+}-m}{2 \pi\left(r_{+}^2-2dr_+-k^2+a^2\right)},
\end{aligned}
\end{equation}
\begin{equation}
\begin{aligned}
S_{B H} & =\pi\left(r_{+}^2-2dr_+-k^2+a^2\right).
\end{aligned}
\end{equation}
Next, we investigate the topological numbers of the dyonic Kerr-Sen BH via R\'{e}nyi statistics. The definition of the generalized free energy is employed to obtain
\begin{equation}
\mathcal{F} _R=\frac{k^2-a^2-p^2-q^2+2dr_+-r_+^2}{2(d-r_+)}-\frac{\mathrm {ln}(1+\lambda \pi (a^2-k^2-2dr_++r_+^2))   }{\lambda \tau _R}.
\end{equation}
The following equation provides the zero points of the component $\phi^{r_+}$
\begin{equation}
\phi^{r_+}=1-\frac{a^2-k^2+p^2+q^2-2dr_++r_+^2}{2(d-r_+)^2} +\frac{2\pi (d-r)}{\tau _R+\lambda \pi(a^2-k^2-2dr_++r_+^2)\tau _R},
\end{equation}
\begin{equation}
\phi ^\theta =-\mathrm { cot} \Theta \mathrm {csc} \Theta,
\end{equation}
thus
\begin{equation}
 \tau _R=\frac{4\pi (d-r_+)^3}{(a^2-2d^2-k^2+p^2+q^2+2dr_+-r_+^2)(1+\pi \lambda (a^2-k^2-2dr_++r_+^2))}  .
\end{equation}
\begin{figure}[htbp]
	\centering
	\subfigure[$\lambda=0$]{
    \begin{minipage}[t]{0.31\linewidth}
		\centering
		\includegraphics[width=2.1in,height=1.8in]{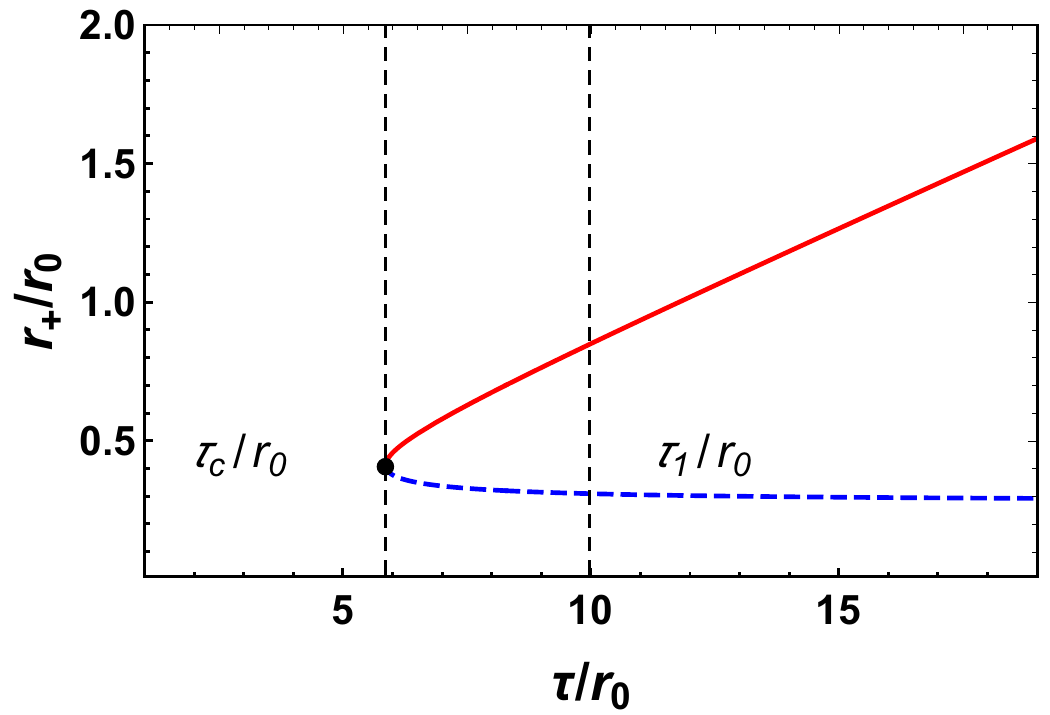}
		\end{minipage}%
            }%
    \subfigure[$\lambda<\lambda_c$]{
    \begin{minipage}[t]{0.31\linewidth}
		\centering
		\includegraphics[width=2.1in,height=1.8in]{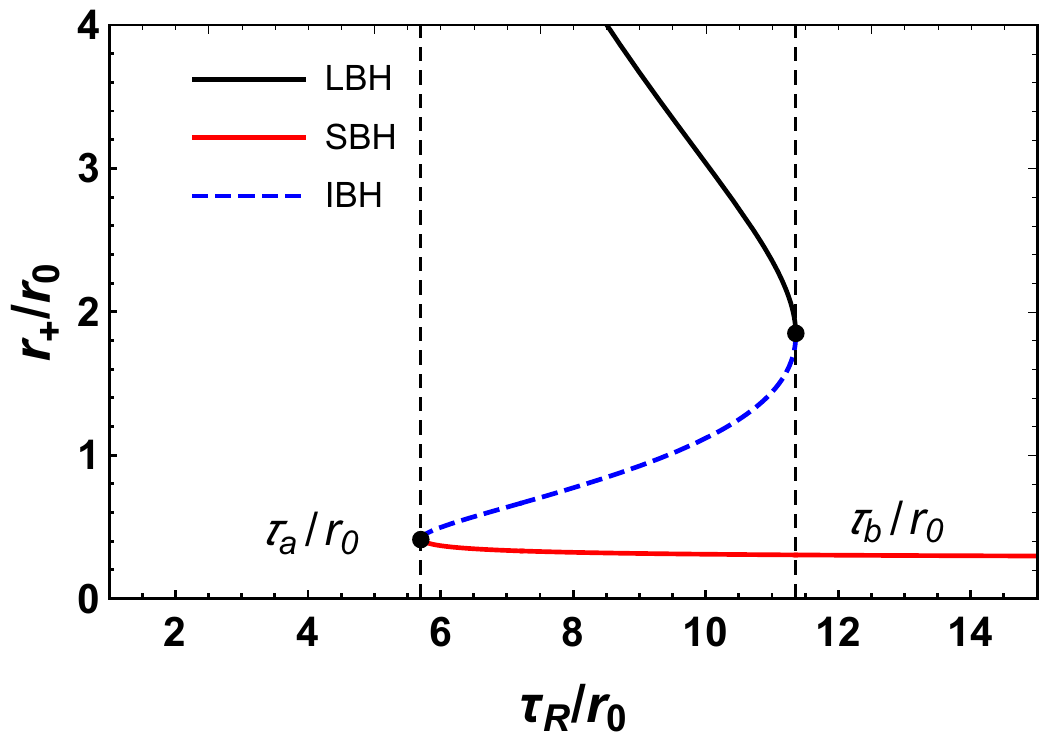}
		\end{minipage}%
            }%
      \subfigure[$\lambda>\lambda_c$]{
    \begin{minipage}[t]{0.31\linewidth}
		\centering
		\includegraphics[width=2.1in,height=1.8in]{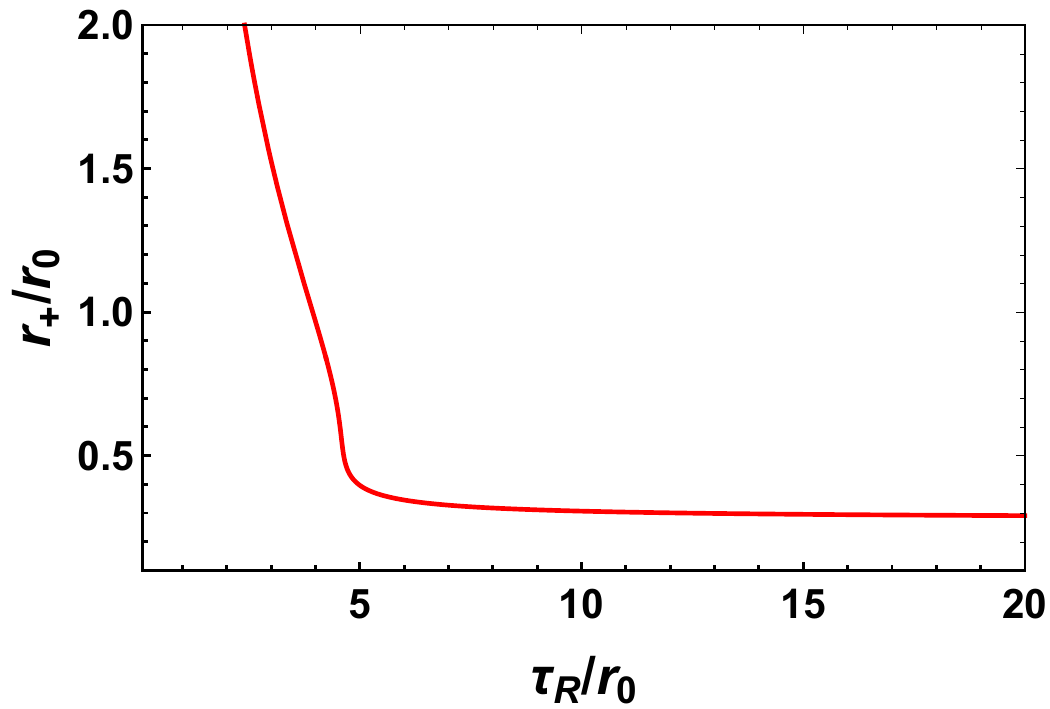}
		\end{minipage}%
            }%
     \centering
     \caption{The zero points of $\phi^{r_+}$ in the $\tau_R-r_+$ plane for dyonic Kerr-Sen BH via R\'{e}nyi statistics.  The left figure (a) is plotted with $\lambda=0$.  The middle figure (b) is plotted with $\lambda r_0^2=0.1$. The right figure (c) is plotted with $\lambda r_0^2=0.8$.}
\end{figure}

\begin{figure}[htbp]
	\centering
	\subfigure[$\lambda=0$]{
    \begin{minipage}[t]{0.31\linewidth}
		\centering
		\includegraphics[width=2.1in,height=1.8in]{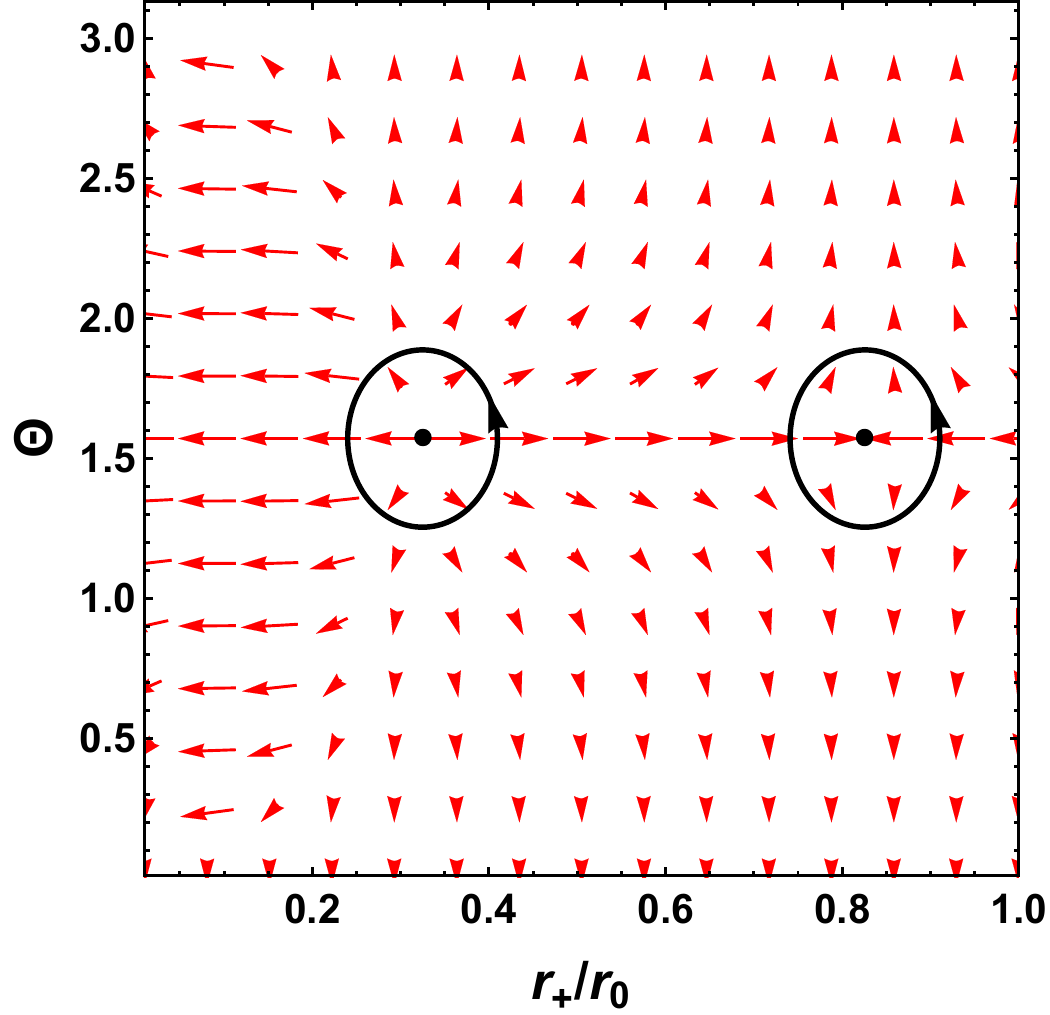}
		\end{minipage}%
            }%
    \subfigure[$\lambda<\lambda_c$]{
    \begin{minipage}[t]{0.31\linewidth}
		\centering
		\includegraphics[width=2.1in,height=1.8in]{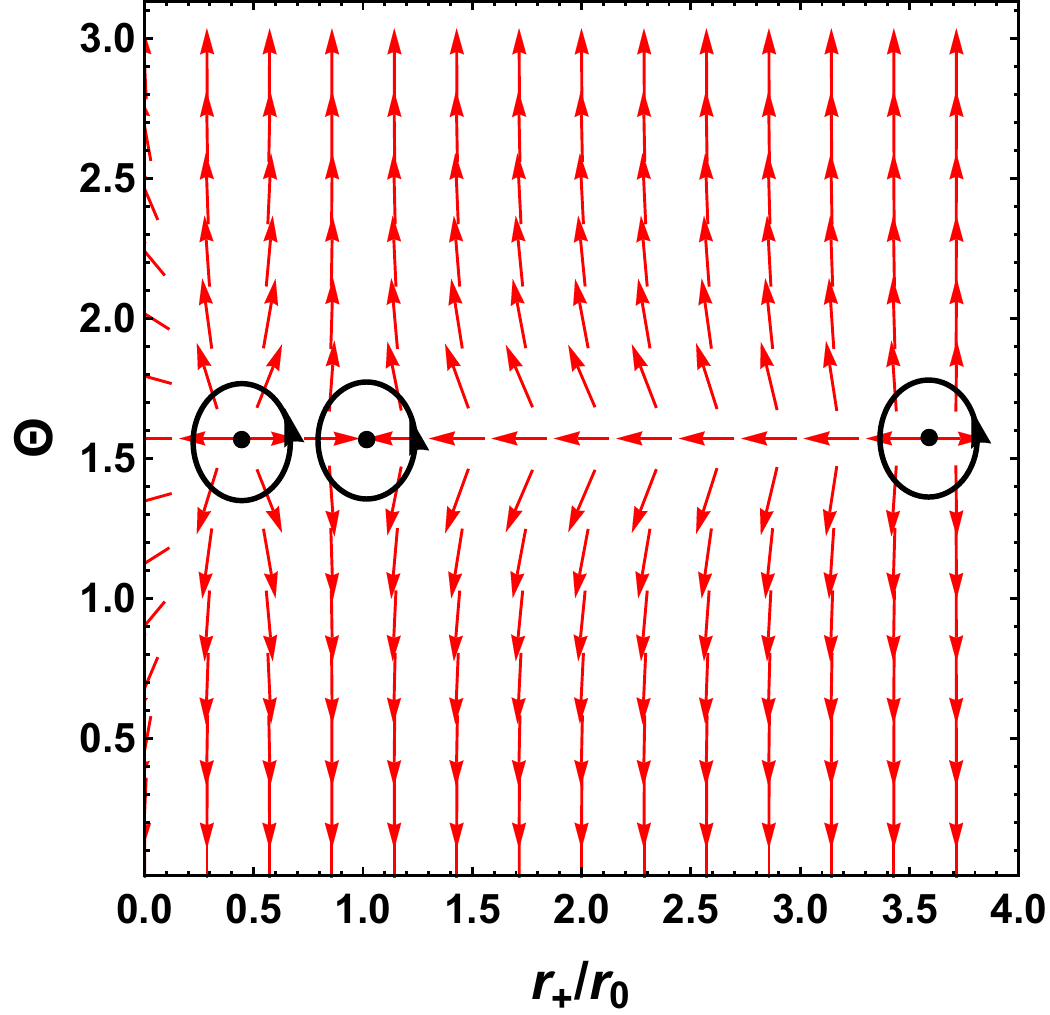}
		\end{minipage}%
            }%
      \subfigure[$\lambda>\lambda_c$]{
    \begin{minipage}[t]{0.31\linewidth}
		\centering
		\includegraphics[width=2.1in,height=1.8in]{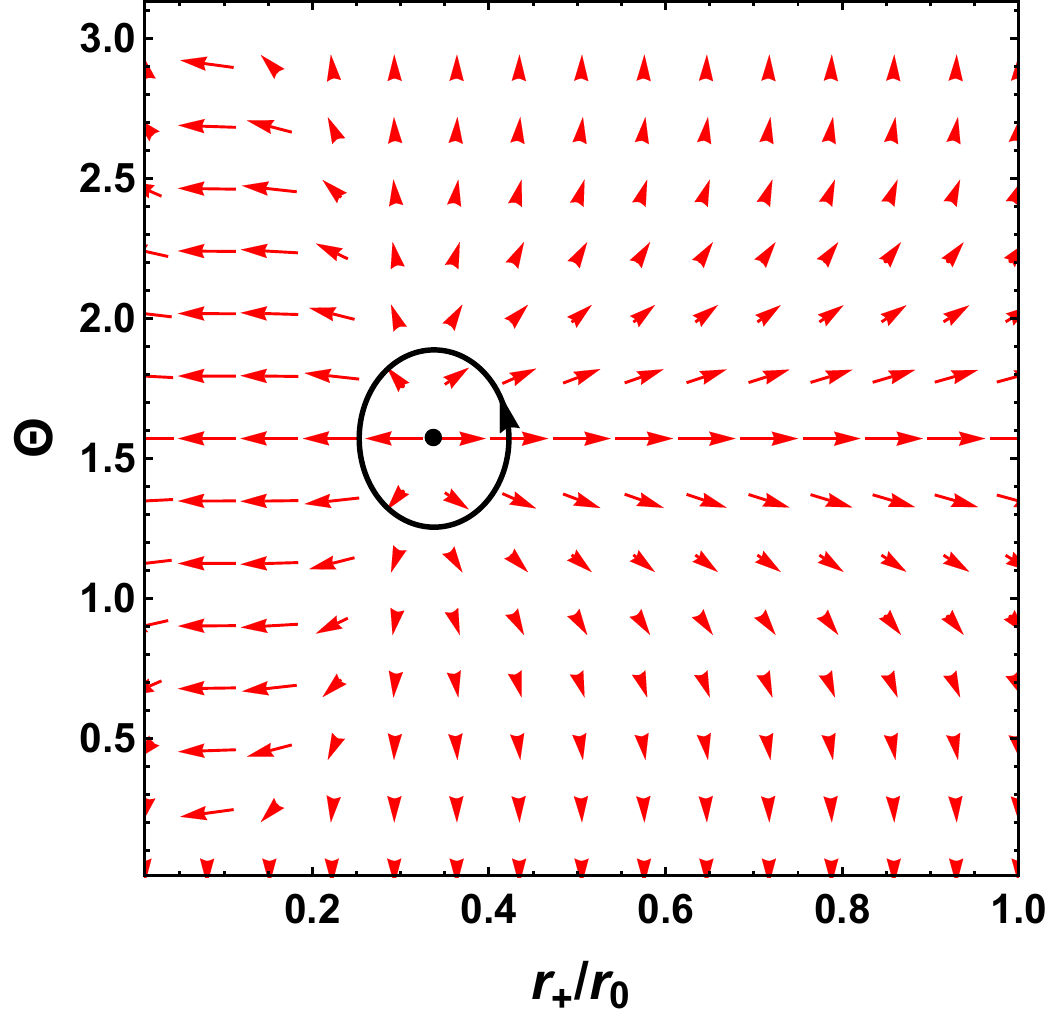}
		\end{minipage}%
            }%
     \centering
     \caption{The unit vector field $n$ on the $r_+-\Theta$ plane for dyonic Kerr-Sen BH via R\'{e}nyi statistics.  The left figure (a) is plotted with $\lambda=0$.  The middle figure (b) is plotted with $\lambda r_0^2=0.1$. The right figure (c) is plotted with $\lambda r_0^2=0.8$.}
\end{figure}
Similar to the analysis in section II. For $\lambda=0$ (the resulting $r_+$ vs $\tau$ as show in FIG. 4(a)), when we set $a/r_0=0.1$, $d/r_0=0.1$, $p/r_0=0.2$ and $q/r_0=0.1$, one have $\tau_c/r_0=5.8606$, when $ \tau >\tau_c$ with $ \tau/r_0=10$, as show in FIG. 5(a), the zero points of $\phi^{r_+}$ are located at $(r_+/r_0, \Theta)=(0.31,\pi/2)$ and $(r_+/r_0, \Theta)=(0.85,\pi/2)$ and  the corresponding topological number $w_1=1$, $w_2=-1$, the total topological number of the dyonic Kerr-Sen BH via GB statistics is $W =0$.
From FIG. 4(b), when $\lambda<\lambda_c$ ($\lambda_c=0.68$ is calculated from equation \eqref{lc}), one have $\tau_a/r_0=5.7153$ and $\tau_b/r_0=11.3595$.  The unit vector field $n$ at $\tau/r_0=9$ is plotted in the FIG. 5(b). The zero points from left to right are at $(r_+/r_0, \Theta)=(0.31,\pi/2)$, $(r_+/r_0, \Theta)=(0.92,\pi/2)$ and $(r_+/r_0, \Theta)=(3.67,\pi/2)$ and corresponding topological number $w_1=1$, $w_2=-1$ and $w_2=1$, so $W =1$. For $\lambda>\lambda_c$, we can see from FIG. 4(c) and FIG. 5(c), we have $W =1$. Thus, the Kerr-Sen BH and dyonic Kerr-Sen BH are in the same topological class under both R\'{e}nyi and GB statistics in this case.

Now let's change at least one of the parameters, such as we set $a/r_0=0.1$, $d/r_0=0.2$, $p/r_0=0.2$ and $q/r_0=0.1$. When $\lambda=0$, from FIG. 6(a) one find there is an unstable BH branch, FIG. 7(a) shows the topological number $W=-1$. When we fix the parameter $\lambda=0.01$, FIG. 6(b) shows that there is an unstable BH branch (dashed blue line) and a stable BH branch (solid red line), they have topological numbers $w_1=-1$ and $w_1=1$, respectively. A annihilation point is located at $ \tau_c/r_0=35.4486$. The unit vector field $n$ at $\tau/r_0=34$ is plotted in the FIG. 7(b). The zero points are $(r_+/r_0, \Theta)=(4.42,\pi/2)$ and $(r_+/r_0, \Theta)=(7.75,\pi/2)$. Thus, the topological number of the dyonic Kerr-Sen BH via R\'{e}nyi statistics is $W =0$. In such parameter space, the Kerr-Sen BH and dyonic Kerr-Sen BH belong to different topological class with both R\'{e}nyi and GB statistics.
\begin{figure}[htbp]
	\centering
    \subfigure[$\lambda=0$]{
    \begin{minipage}[t]{0.4\linewidth}
		\centering
		\includegraphics[width=2.5in,height=2.0in]{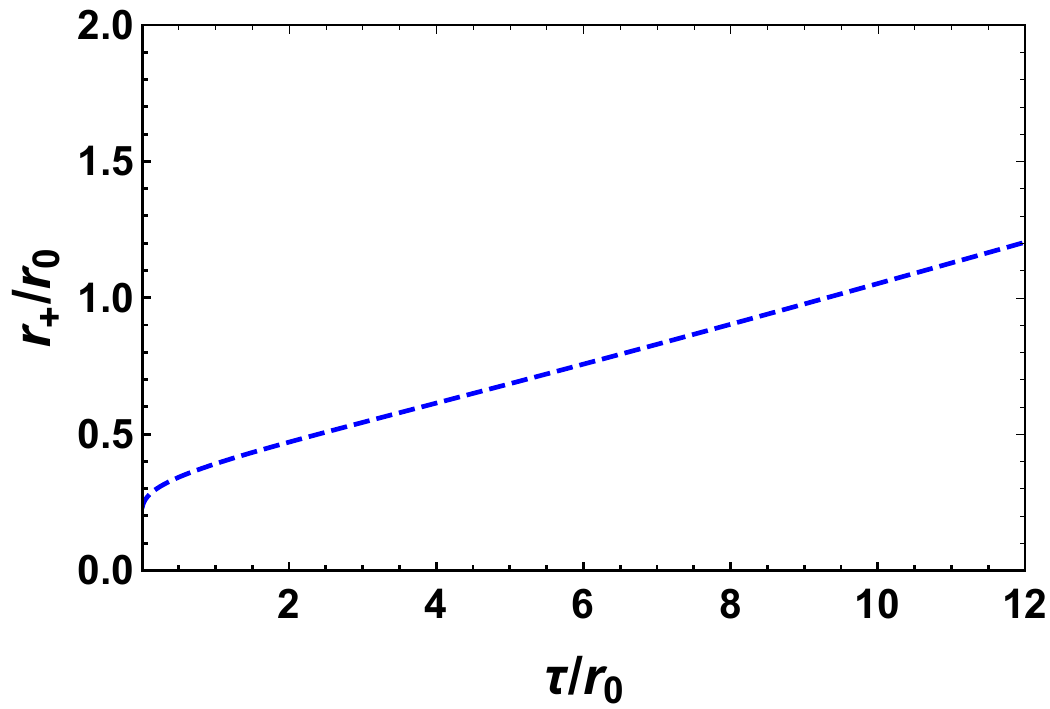}
		\end{minipage}%
            }%
      \subfigure[$\lambda=0.01$]{
    \begin{minipage}[t]{0.4\linewidth}
		\centering
		\includegraphics[width=2.5in,height=2.0in]{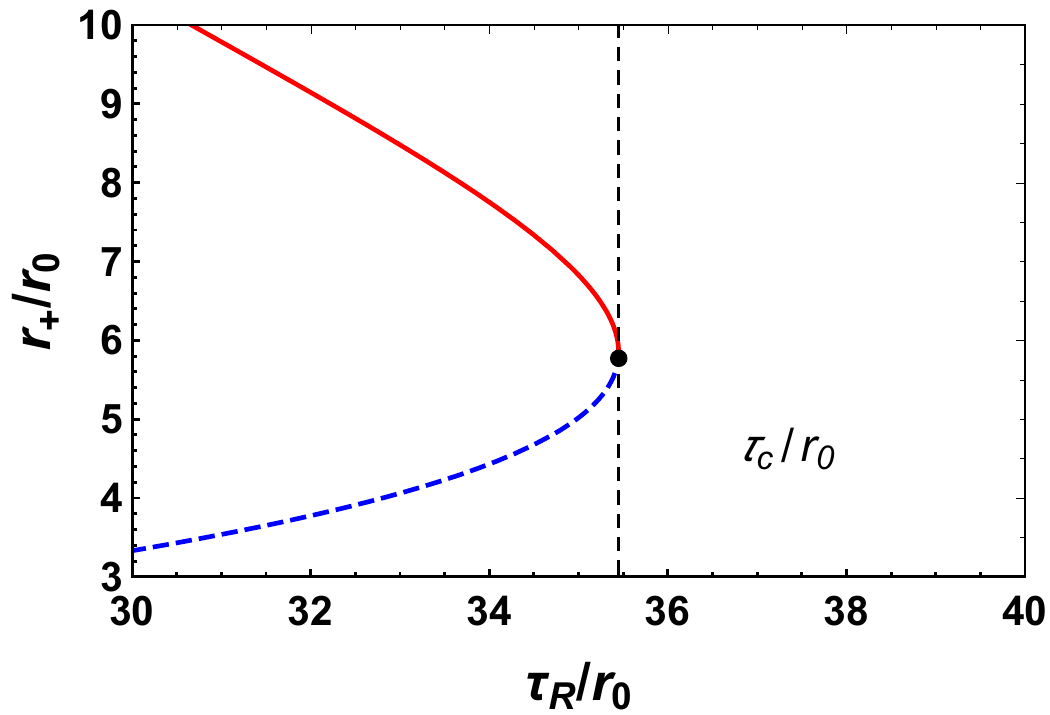}
		\end{minipage}%
            }%
     \centering
     \caption{The zero points of $\phi^{r_+}$ in the $\tau_R-r_+$ plane for dyonic Kerr-Sen BH via R\'{e}nyi statistics.  The left figure (a) is plotted with $\lambda=0$.  The right figure (b) is plotted with $\lambda r_0^2=0.01$. }
\end{figure}

\begin{figure}[htbp]
	\centering
    \subfigure[$\lambda=0$]{
    \begin{minipage}[t]{0.4\linewidth}
		\centering
		\includegraphics[width=2.5in,height=2.0in]{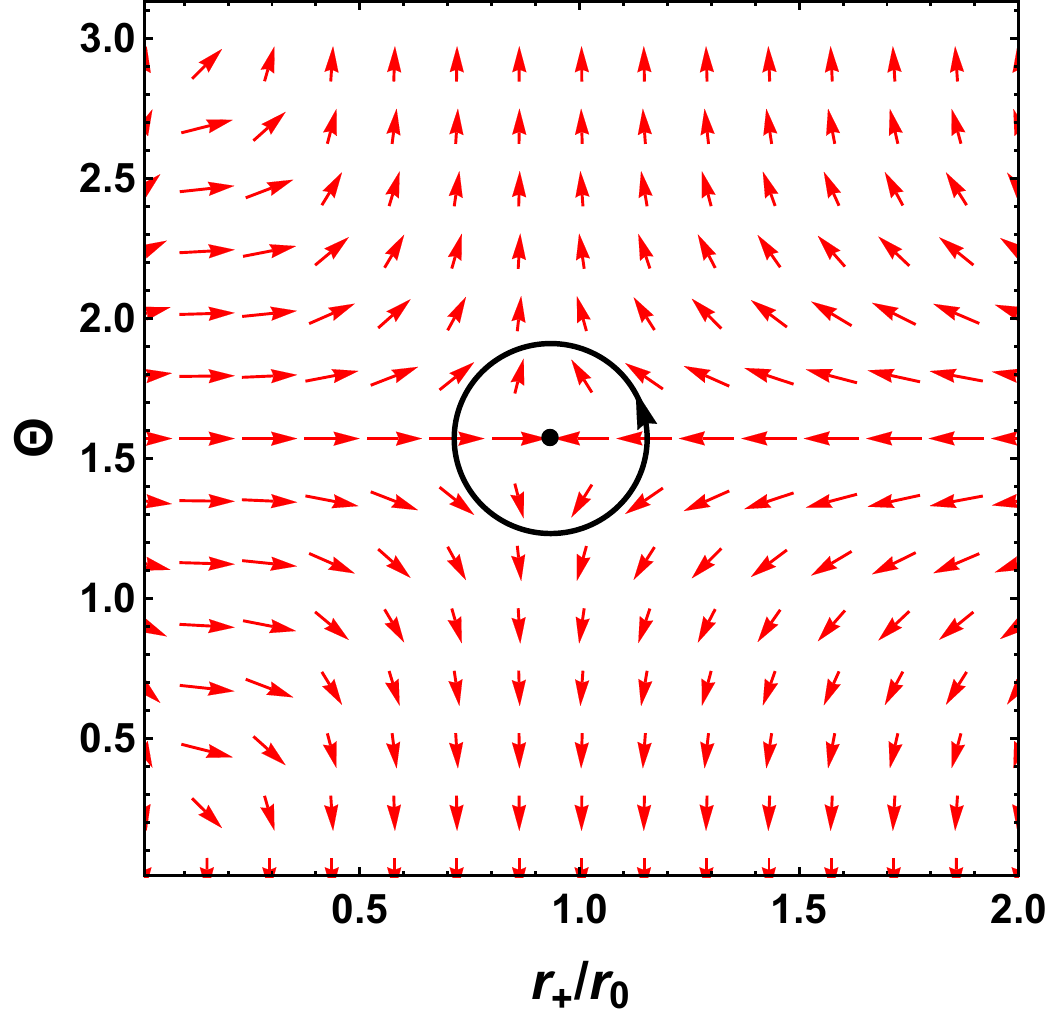}
		\end{minipage}%
            }%
      \subfigure[$\lambda=0.01$]{
    \begin{minipage}[t]{0.4\linewidth}
		\centering
		\includegraphics[width=2.5in,height=2.0in]{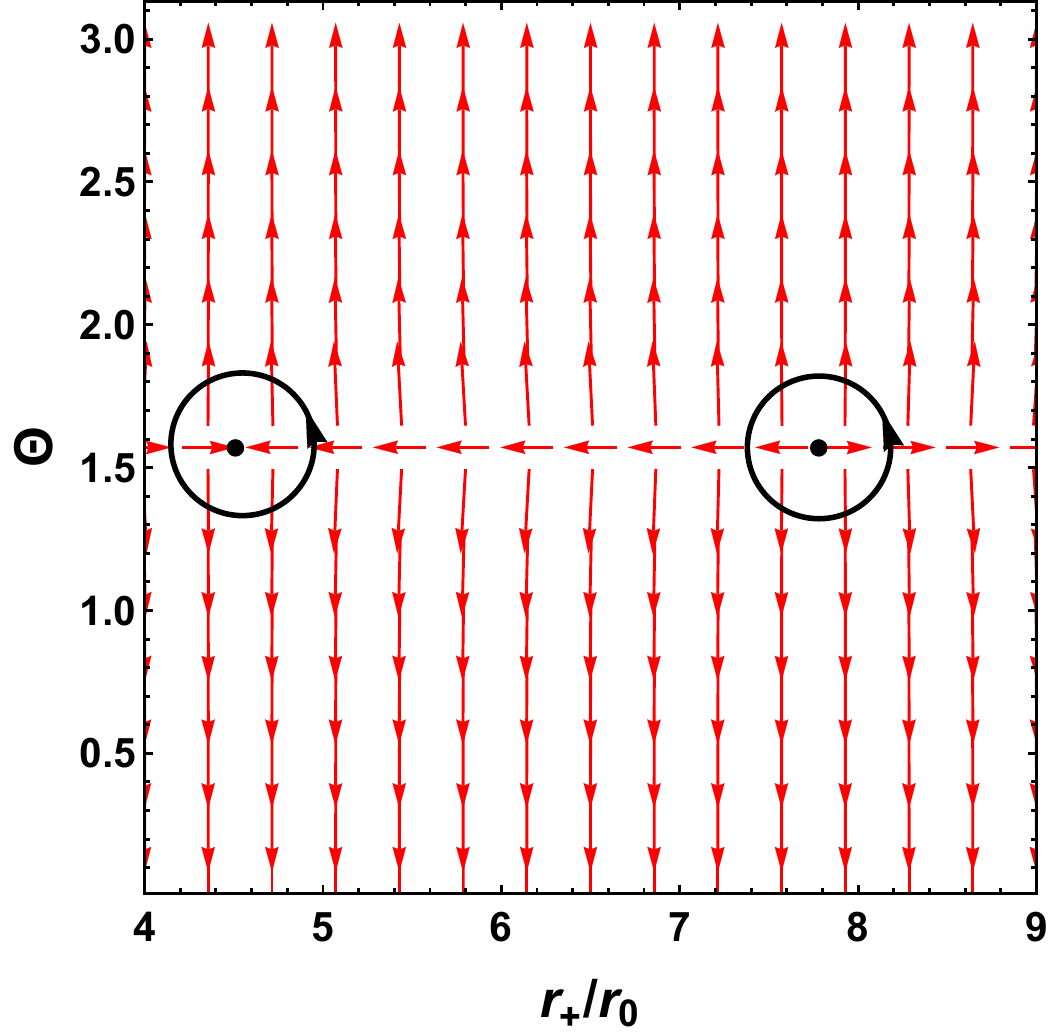}
		\end{minipage}%
            }%
     \centering
     \caption{The unit vector field $n$ on the $r_+-\Theta$ plane for dyonic Kerr-Sen BH via R\'{e}nyi statistics.  The left figure (a) is plotted with $\lambda=0$.  The right figure (b) is plotted with $\lambda r_0^2=0.01$. }
\end{figure}
\section{Dyonic Kerr-Sen-AdS BH via GB statistics}
We extend the dyonic Kerr-Sen solution in previous section to the AdS case with GB statistics. The metric for the dyonic Kerr-Sen-AdS BH has the form \cite{K14}
\begin{equation}
\begin{aligned}
d \bar{s}^2=-\frac{\bar{\Delta}_r}{\bar{\Sigma}} \bar{X}^2+\frac{\bar{\Sigma}}{\bar{\Delta}_r} d r^2+\frac{\bar{\Sigma}}{\bar{\Delta}_\theta} d \theta^2+\frac{\bar{\Delta}_\theta \sin ^2 \theta}{\bar{\Sigma}} \bar{Y},
\end{aligned}
\end{equation}
where
\begin{equation}
\begin{aligned}
\bar{X}= & d t-\frac{a \sin ^2 \theta}{\Xi} d \bar{\varphi}, \quad \bar{Y}=a d t-\frac{r^2-2 d r-k^2+a^2}{\Xi} d \bar{\varphi}, \\
\bar{\Delta}_r= & \left(1+\frac{r^2-2 d r-k^2}{l^2}\right)\left(r^2-2 d r-k^2+a^2\right) -2 m(r-d)+p^2+q^2,\\
\bar{\Delta}_\theta= & 1-\frac{a^2}{l^2} \cos ^2 \theta, \quad \Xi=1-\frac{a^2}{l^2} .
\end{aligned}
\end{equation}
The ADM mass $M$, and other thermodynamic quantities in AdS spacetimes are related as
\begin{equation}
\begin{aligned}
M & =\frac{m}{\Xi},\\
T_{BH} & =\frac{\left(r_{+}-d\right)\left(2 r_{+}^2-4 d r_{+}-2 k^2+a^2+l^2\right)-m l^2}{2 \pi\left(r_{+}^2-2 d r_{+}-k^2+a^2\right) l^2},\\
S_{B H} & =\frac{\pi}{\Xi}\left(r_{+}^2-2 d r_{+}-k^2+a^2\right), \\
\end{aligned}
\end{equation}
and we use the following generalized free energy
\begin{equation}
\mathcal{F} =\frac{3p^2+3q^2-\left(a^2-k^2-2 dr_+ + r_+^2\right)\left(8\pi P (k^2+2 d r_+ - r_+^2)-3\right)}{2\left(8\pi P a^2 -3\right)(d-r_+)}+\frac{3 \pi\left(a^2-k^2-2dr_+ +r^2_+ \right)}{\left(8 \pi P a^2-3\right) \tau}.
\end{equation}
\begin{figure}[htbp]
	\centering
    \subfigure[]{
    \begin{minipage}[t]{0.4\linewidth}
		\centering
		\includegraphics[width=2.5in,height=2.0in]{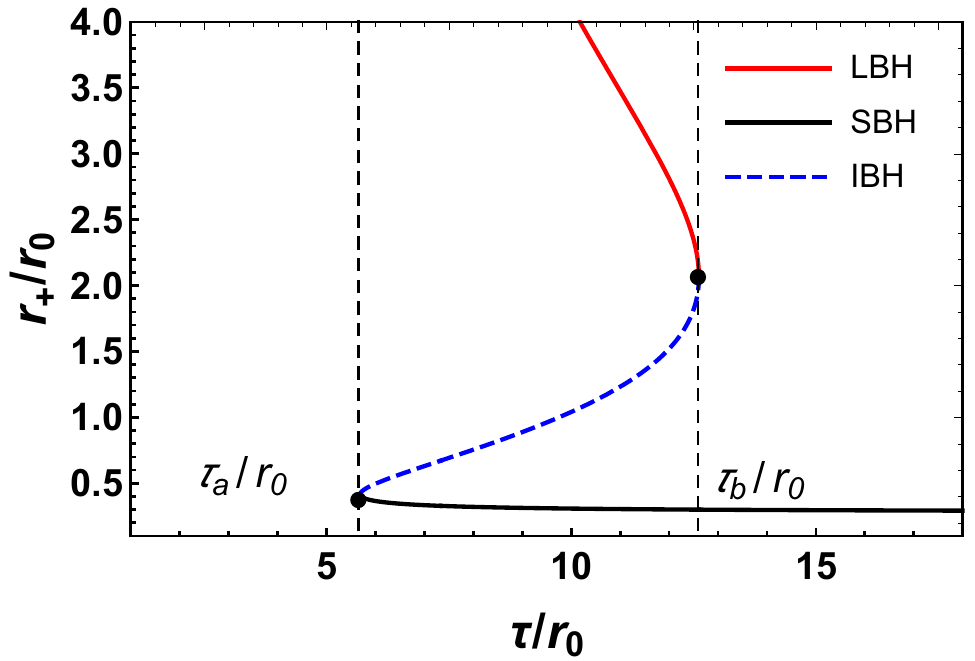}
		\end{minipage}%
            }%
      \subfigure[]{
    \begin{minipage}[t]{0.4\linewidth}
		\centering
		\includegraphics[width=2.5in,height=2.0in]{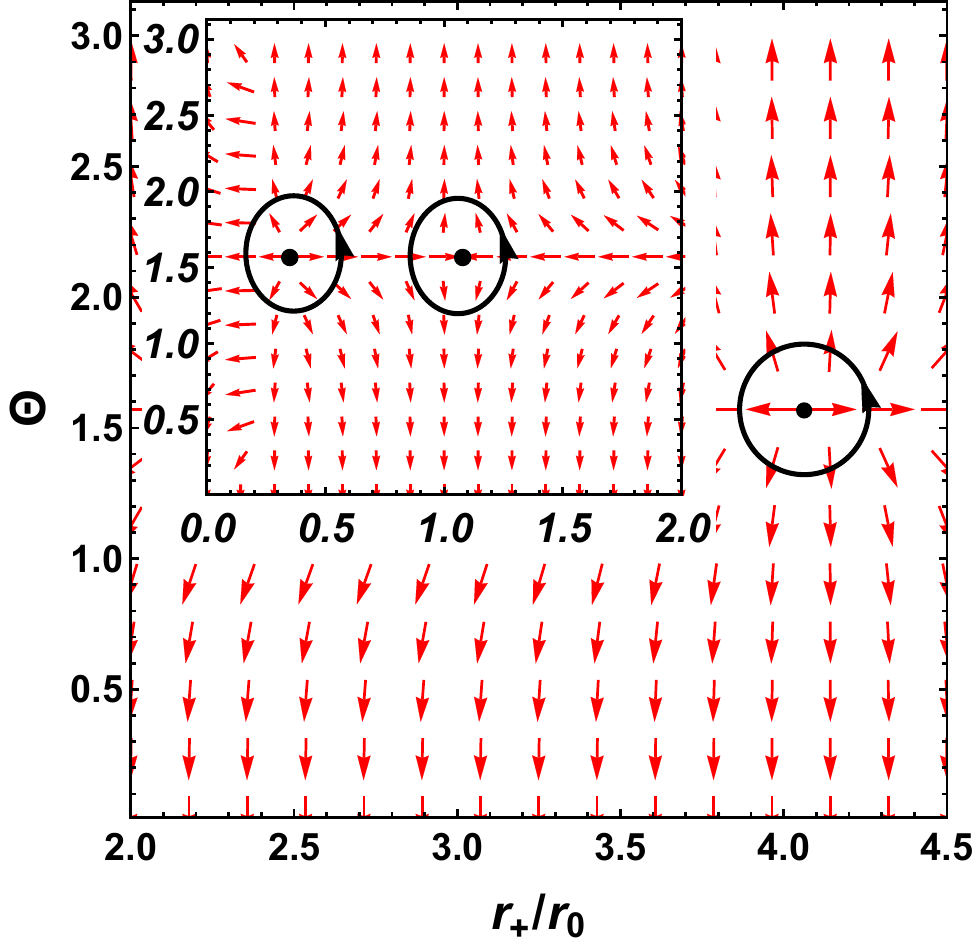}
		\end{minipage}%
            }%
     \centering
     \caption{Topological properties for dyonic Kerr-Sen-AdS BH via GB statistics. The left figure (a) reprensents of the zero points $\phi^{r_+}$ in $\tau-r_+$ plane. The right figure (b) reprensents the unit vector field $n$ on the $r_+-\Theta$ plane with $\tau/r_0=10$.}
\end{figure}
Thus, the vector components can be written as
\begin{equation}
\phi^{r_+}=\frac{\mathcal{A}_{r_+}\tau-12 \pi(d-r_+)^3}{2\left(8 \pi P a^2 -3\right)(d-r_+)^2 \tau},
\end{equation}
where
\begin{equation}
\begin{aligned}
\mathcal{A}_{r_+}=&3\left(p^2+q^2-k^2\right)+64 \pi P d^3 r_+-3 r_+^2+8\pi P \left(k^4+2 k^2 r_+^2-3 r_+^4\right)+2 d^2\left(16 \pi P\left(k^2-4 r_+^2\right)-3\right)\\
&+2 d r_+\left(3-16 \pi P \left(k^2-3 r_+^2\right)\right)+a^2\left(3-8\pi P \left(2 d^2+k^2-2 d r_++r_+^2\right)\right),
\end{aligned}
\end{equation}
and
\begin{equation}
\phi ^\theta =-\mathrm { cot} \Theta \mathrm {csc} \Theta,
\end{equation}
so we have
\begin{equation}
\tau =\frac{12 \pi(d-r_+)^3}{\mathcal{A}_{r_+}}.
\end{equation}
For $a/r_0=0.1$, $d/r_0=0.1$, $p/r_0=0.2$, $q/r_0=0.1$ and $Pr_0^2=0.01 (P<P_c)$, from FIG. 8, we plot the $r_+$ vs $\tau$ graph and the unit vector field $n$. Similar to the Kerr-Sen-AdS BH via GB statistics, for $\tau_a<\tau<\tau_b$, the stable branches of large and small black holes are represented by solid red and black lines, the zero points on these branches having a topological number of $w=1$, the blue dotted curve represents an unstable intermediate BH branch, with the topological number of the zero point on that branch being $w=-1$. Thus, the topological number of  dyonic Kerr-Sen-AdS BH is $W=1+1-1=1$. The generation and annihilation points at $\tau_a/r_0=5.6876$ and $\tau_b/r_0=12.6096$, respectively. The zero points in FIG. 8(b) from left to right are at $(r_+/r_0, \Theta)=(0.31,\pi/2)$, $(r_+/r_0, \Theta)=(1.04,\pi/2)$ and $(r_+/r_0, \Theta)=(4.11,\pi/2)$ with $\tau/r_0=10$. Such a parameter space implies that the topological numbers of the dyonic Kerr-Sen-AdS BH and Kerr-Sen-AdS BH are the same.

Similarly, if we change the parameter space, i.e, $a/r_0=0.1$, $d/r_0=0.2$, $p/r_0=0.2$ and $q/r_0=0.1$ and $Pr_0^2=0.1$. From FIG. 9 we find the curves of $r_+$ vs $\tau$ look similar to the case of dyonic Kerr-Sen BH via
the R\'{e}nyi statistics with $\lambda =0.01$. We can obtain the topological number $W$ for the case is equal to 0, which is different from the case of Kerr-Sen-AdS.
\begin{figure}[htbp]
	\centering
    \subfigure[]{
    \begin{minipage}[t]{0.4\linewidth}
		\centering
		\includegraphics[width=2.5in,height=2.0in]{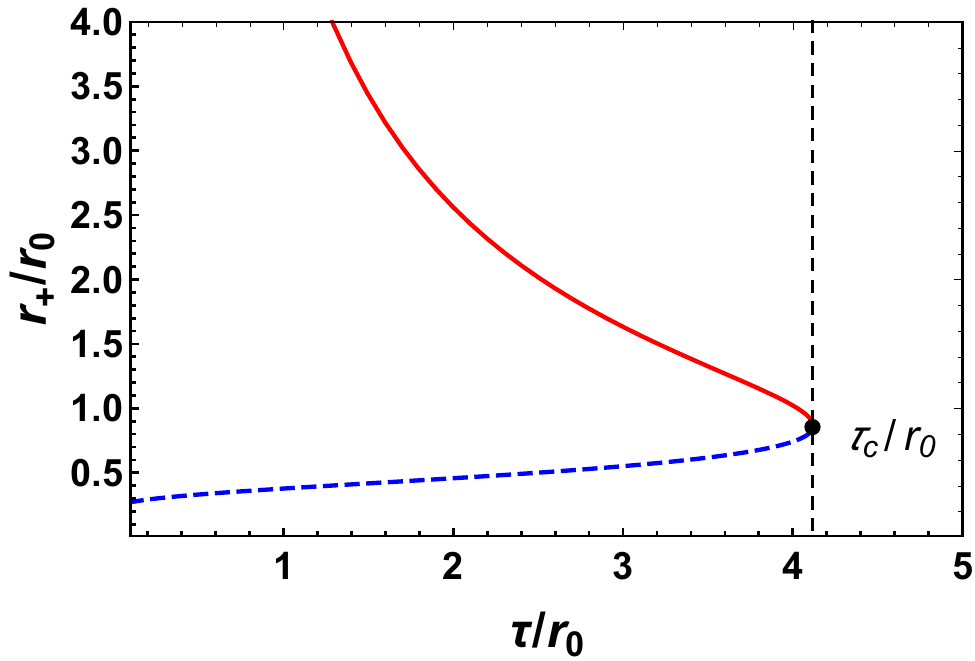}
		\end{minipage}%
            }%
      \subfigure[]{
    \begin{minipage}[t]{0.4\linewidth}
		\centering
		\includegraphics[width=2.5in,height=2.0in]{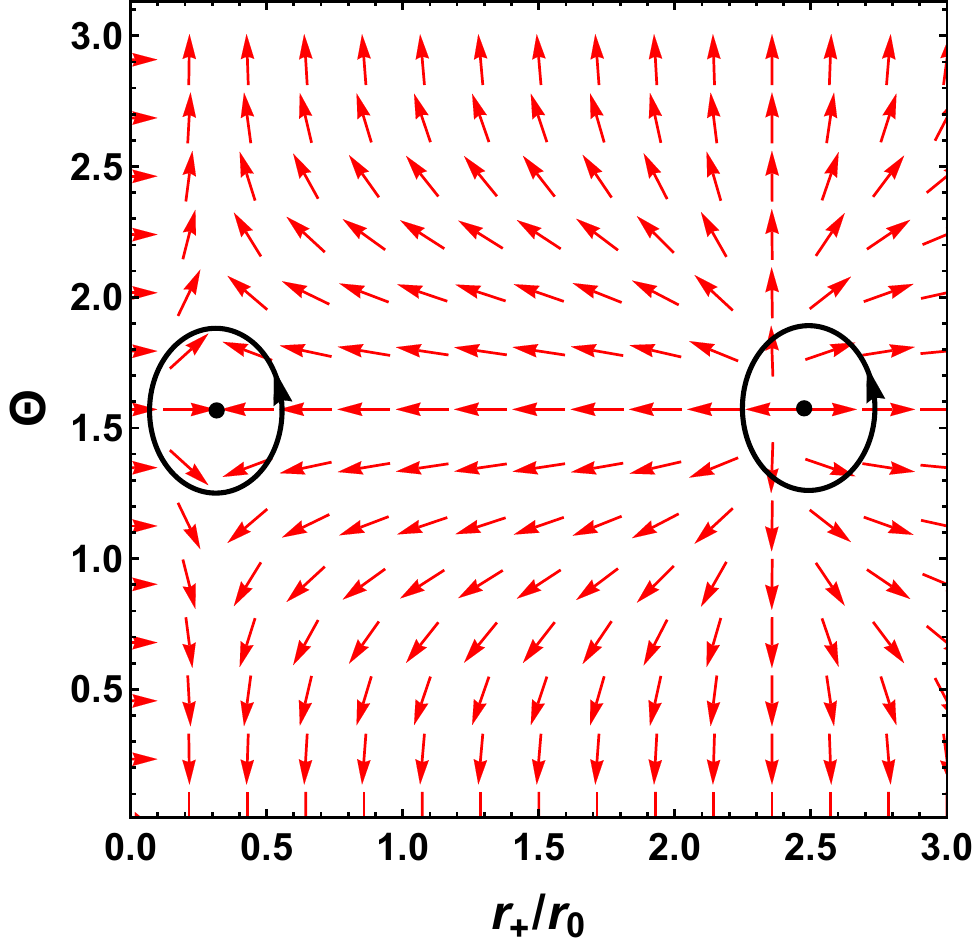}
		\end{minipage}%
            }%
     \centering
     \caption{Topological properties for dyonic Kerr-Sen-AdS BH via GB statistics. The left figure (a) reprensents the  zero points of $\phi^{r_+}$ in $\tau$ vs $r_+$ plane. The right figure (b) reprensents the unit vector field $n$ on the $r_+-\Theta$ plane with $\tau/r_0=2$.}
\end{figure}

\begin{table}[H]
	\centering
	\caption{In following table, we summarize the topological numbers associated with Kerr-Sen and dyonic Kerr-Sen black holes by using R\'{e}nyi entropy processes, are well as the AdS cases of such two black holes via GB entropy, the annihilation point (AP) and generation point (GP) of black holes are also given.}
\begin{tabular}{c|p{2.5cm}<{\centering}|p{3cm}<{\centering}|p{1cm}<{\centering}|p{1cm}<{\centering}|p{1cm}<{\centering}|p{1cm}<{\centering}}
 \hline \hline\multirow{2}{*}{Black holes (BHs)} &\multirow{2}{*}{GB statistics~(W)}    & \multirow{2}{*}{R\'{e}nyi statistics~(W)} & \multicolumn{2}{|l}{~~GB statistics }  &\multicolumn{2}{|l}{R\'{e}nyi statistics}  \\\cline{4-7}
              &                                         &                                           & GP                      &  AP                     & GP                      &  AP           \\
 \hline      Kerr-Sen BH                             & 0                                            & 1      &1       &0       & 1~or~0 &   1~or~0\\
       dyonic Kerr-Sen BH                            &0~or~-1             &   1~or~0                      & 1~or~0                         &   0                     &        1~or~0                 &   1~or~0        \\
\hline       Kerr-Sen-AdS BH                         & 1 & -      &1~or~0 & 1~or~0 &-       &  -\\
       dyonic Kerr-Sen-AdS  BH                       & 1~or~0                    &   -                      &     1~or~0                      &        1~or~0                   &     -                    &     -    \\
\hline \hline
\end{tabular}
\end{table}
\section{CONCLUSIONS}
In this paper, we investigated thermodynamic topology of Kerr-Sen BH and dyonic Kerr-Sen BH in the context of R\'{e}nyi statistics approach. By representing  the Hawking-Bekenstein entropy of BH horizons as a non-extensive R\'{e}nyi entropy, one obtained generalized off-shell free energy on the R\'{e}nyi entropy. The findings of our study as presented in this paper are now summarized in the TABLE I. In doing this study, we have taken into consideration the Kerr-Sen BH and dyonic Kerr-Sen BH within the framework of R\'{e}nyi statistics, the corresponding topological number shows different value compared to the conventional GB statistics. It is worth noting that the non-extensive parameter $\lambda$, as an additional parameter, changes the topological number. Moreover, the topological classifications of Kerr-Sen and dyonic Kerr-Sen black holes remain consistent in both GB and R\'{e}nyi statistics. Indeed, we have demonstrated that there are places in the parameter space where Kerr-Sen and dyonic Kerr-Sen black holes belong to the same (different) topology class in both GB and R\'{e}nyi statistics.

Furthermore, we calculated the topological numbers of Kerr-Sen BH and dyonic Kerr-Sen BH in the presence of the cosmological constant $\Lambda$ with GB statistics. The topological numbers of Kerr-Sen-AdS and dyonic Kerr-Sen-AdS black holes by using GB entropy were found to be consistent with these two types of black holes in the absence of cosmological constant with R\'{e}nyi entropy processes. Such an analogy suggests a significant link between the nonextensive R\'{e}nyi parameter $\lambda$ and the cosmological constant $\Lambda$ from topology perspective.

One intriguing question is other black holes solutions and high-dimensional space-time also consider such an analysis in \cite{T26}. There is no doubt that, the above  mentioned equivalence require more scrutiny in other systems, possibly involving reentrant phase transitions and the triple point phenomenon. It may be worthwhile to explore the application of R\'{e}nyi statistics in thermodynamic topology and investigate potential deviations from conventional GB statistics, which could provide insights into the nature of correlations within the system. Additionally, the topological analysis framework based on the nonextensive R\'{e}nyi formalism is an essential approach for comprehending the nontrivial thermodynamical behavior from a thermodynamic topology perspective. We hope that issues beyond the present report will be addressed in future work.
\section*{Acknowledgments}
The authors would like to thank the referee for his/her valuable comments and suggestions that improved the paper. We are very grateful to Di Wu for helpful correspondence. This work is supported by the National Natural Science Foundation of China (Grant No.12265007) and the Doctoral Foundation of Zunyi Normal University of China (BS [2022] 07, QJJ-[2022]-314). The research was partially supported by the Long-Term Conceptual Development of a University of Hradec Kr\'{a}lov\'{e} for 2023, issued by the Ministry of Education, Youth, and Sports of the Czech Republic.





\begin{thebibliography}{0}
\expandafter\ifx\csname natexlab\endcsname\relax\def\natexlab#1{#1}\fi
\expandafter\ifx\csname bibnamefont\endcsname\relax
  \def\bibnamefont#1{#1}\fi
\expandafter\ifx\csname bibfnamefont\endcsname\relax
  \def\bibfnamefont#1{#1}\fi
\expandafter\ifx\csname citenamefont\endcsname\relax
  \def\citenamefont#1{#1}\fi
\expandafter\ifx\csname url\endcsname\relax
  \def\url#1{\texttt{#1}}\fi
\expandafter\ifx\csname urlprefix\endcsname\relax\def\urlprefix{URL }\fi
\providecommand{\bibinfo}[2]{#2}
\providecommand{\eprint}[2][]{\url{#2}}

\end{thebibliography}


\begin{thebibliography}{99}
\bibitem{K1} A. M. Ghezelbash and H. M. Siahaan, Class. Quant. Grav. 30, 135005 (2013).
\bibitem{K2} M. Rogatko, Phys. Rev. D 82, 044017 (2010).
\bibitem{K3} J. Jiang, X. Liu and M. Zhang, Phys. Rev. D 100, 084059 (2019).
\bibitem{K4} H. M. Siahaan, Phys. Rev. D 93, 064028 (2016).
\bibitem{K5} S. V. M. C. B. Xavier, P. V. P. Cunha, L. C. B. Crispinoand C. A. R. Herdeiro, Int. J. Mod. Phys. D 29, 2041005 (2020).
\bibitem{K6} Z. Younsi, A. Zhidenko, L. Rezzolla, R. Konoplya and Y. Mizuno, Phys. Rev. D 94, 084025 (2016).
\bibitem{K7} J.-i. Koga and K.-i. Maeda, Phys. Rev. D 52, 7066 (1995).
\bibitem{K8} K. Hioki and U. Miyamoto, Phys. Rev. D 78, 044007 (2008).
\bibitem{K9} J. H. Horne and G. T. Horowitz, Phys. Rev. D 46, 1340 (1992).
\bibitem{K10} H. Furuhashi and Y. Nambu, Prog. Theor. Phys. 112, 983 (2004).
\bibitem{K11} H. M. Siahaan, Int. J. Mod. Phys. D 24, 1550102 (2015).
\bibitem{K12} M. F. A. R. Sakti, Eur. Phys. J. C 83, 255 (2023).
\bibitem{K13} M. F. A. R. Sakti and P. Burikham, Phys. Rev. D 106, 106006 (2022).
\bibitem{K14} D. Wu, S. Q. Wu, P. Wu, and H. Yu, Phys.Rev. D 103, 044014 (2021).
\bibitem{K15} S. Jana, S. Kar, Phys. Rev. D 108, 044008 (2023).
\bibitem{K16} H. L. Prihadi, F. P. Zen, D. Dwiputra and S. Ariwahjoedi, Phys. Rev. D 107, 124053 (2023).

\bibitem{R1} S. W. Hawking, D. N. Page, Commun. Math. Phys. 87, 577 (1983).
\bibitem{R2} J. D. Bekenstein, Phys. Rev. D 7, 2333 (1973).
\bibitem{R3} L. Y. Hung, R. C. Myers, M. Smolkin, and A. Yale, JHEP 12, 047 (2011).
\bibitem{R4} A. Giveon and D. Kutasov, JHEP 01, 042 (2016).
\bibitem{R5} X. Dong, Nature Commun. 7, 12472 (2016).
\bibitem{R6} M. Crossley, E. Dyer, and J. Sonner, JHEP 12, 001 (2014).
\bibitem{R7} S. D. Haro, J. v. Dongen, M. Visser, and J. Butterfield, Stud. Hist. Phil. Sci. B 69, 82 (2020).
\bibitem{R8} C. Tsallis, Entropy 22, 17 (2019).
\bibitem{R9} A. Alonso-Serrano, M. P. D. abrowski, and H. Gohar, Phys. Rev. D 103, 026021 (2021),
\bibitem{R10} M. Headrick, A. Maloney, E. Perlmutter, and I. G. Zadeh, JHEP 07, 059 (2015).
\bibitem{R11} V. G. Czinner and H. Iguchi, Phys. Lett. B 752, 306 (2016).
\bibitem{R12} V. G. Czinner and H. Iguchi, Eur. Phys. J. C 77, 892 (2017).
\bibitem{R13} L. Tannukij, P. Wongjun, E. Hirunsirisawat, T. Deesuwan and C. Promsiri, Eur. Phys. J. Plus 135, 500 (2020).
\bibitem{R14} C. Promsiri, E. Hirunsirisawat and W. Liewrian, Phys. Rev. D 102, 064014 (2020).
\bibitem{R15} C. Promsiri, E. Hirunsirisawat and W. Liewrian, Phys. Rev. D 104, 064004 (2021).
\bibitem{R16} R. Nakarachinda, E. Hirunsirisawat, L. Tannukij and P. Wongjun, Phys. Rev. D 104, 064003 (2021).
\bibitem{R17} C. Promsiri, E. Hirunsirisawat, and R. Nakarachinda, Phys. Rev. D 105, 124049 (2022).

\bibitem{R18} F. Barzi, H. El Moumni, K. Masmar, Gen. Rel. Grav. 55, 109 (2023).
\bibitem{R19} P. Chunaksorn, E. Hirunsirisawat, R. Nakarachinda, L. Tannukij, P. Wongjun, Eur. Phys. J. C 82, 1174 (2022).
\bibitem{R20} Z. Wang, H. Ren, J. Chen and Y. Wang, Eur. Phys. J. C 83, 527 (2023).
\bibitem{R21} F. Barzi, H. El Moumni, K. Masmar, JHEAp 42, 63 (2024).
\bibitem{T26} C. W. Tong, B. H. Wang and J. R. Sun, arxiv:2310.09602[gr-qc].


\bibitem{T1} S. W. Wei, Y. X. Liu and R. B. Mann, Phys. Rev. Lett. 129, 191101 (2022).
\bibitem{T2} N. C. Bai, L. Li and J. Tao, Phys. Rev. D 107, 064015 (2023).
\bibitem{T3} P. K. Yerra and C. Bhamidipati, Phys. Rev. D 105, 104053 (2022).
\bibitem{T4} P. K. Yerra, C. Bhamidipati and S. Mukherji, Phys. Rev. D 106, 064059 (2022).
\bibitem{T5} P. K. Yerra and C. Bhamidipati, Phys. Lett. B 835, 137591 (2022).
\bibitem{T6} P. K. Yerra, C. Bhamidipati and S. Mukherji, JHEP 03, 138 (2024).
\bibitem{Tx1} P. K. Yerra, C. Bhamidipati, and S. Mukherji, J. Phys. Conf. Ser. 2667, 012031 (2023).
\bibitem{T7} D. Wu and S. Q. Wu, Phys. Rev. D 107, 084002 (2023).
\bibitem{T8} D. Wu, Eur. Phys. J. C 83, 589 (2023).
\bibitem{T9} D. Wu, Phys. Rev. D 108, 084041 (2023).
\bibitem{T10} D. Wu, Eur. Phys. J. C 83, 365 (2023).
\bibitem{T11} D. Wu, Phys. Rev. D 107, 024024 (2023).
\bibitem{T12} Y. B. Du and X.D. Zhang, Eur. Phys. J. C 83, 927 (2023).
\bibitem{T13} Z. Y. Fan, Phys. Rev. D 107, 044026 (2023).
\bibitem{T14} C. X. Fang, J. Jiang and M. Zhang, JHEP 01, 102 (2023).
\bibitem{T15} M. Zhang and J. Jiang, JHEP 06, 115 (2023).
\bibitem{T16} C. Liu and J. Wang, Phys. Rev. D 107, 064023 (2023).
\bibitem{T17} R. Li, C. H. Liu, K. Zhang and J. Wang, Phys. Rev. D 107, 044003 (2023).
\bibitem{T18} M. Y. Zhang, H. Chen, H. Hassanabadi, Z. W. Long and H. Yang, Eur. Phys. J. C 83, 773 (2023).
\bibitem{T19} T. N. Hung and C. H. Nam, Eur. Phys. J. C 83, 582 (2023).
\bibitem{T20} M. Rizwan and K. Jusufi, Eur. Phys. J. C 83, 944 (2023).
\bibitem{T21} N. J. Gogoi and P. Phukon, Phys. Rev. D 107, 106009 (2023).
\bibitem{T22} D. Y. Chen, Y. C. He and J. Tao, Eur. Phys. J. C 83, 872 (2023).
\bibitem{T23} M. R. Alipour, M. A. S. Afshar, S. N. Gashti and J. Sadeghi, Phys. Dark Univ. 42, 101361 (2023).
\bibitem{T24} N. J. Gogoi and P. Phukon, Phys. Rev. D 108, 066016 (2023).
\bibitem{T25} J. Sadeghi, S. N. Gashti, M. R. Alipour and M. A. S. Afshar, Annals Phys. 455, 169391 (2023).
\bibitem{Tx2} J. Sadeghi, M. A. S Afshar, S. N Gashti, M. R. Alipour, Annals Phys. 460, 169569 (2024).
\bibitem{Tx3} J. Sadeghi, M. A. S. Afshar, S. N. Gashti, M. R. Alipour, Astropart. Phys. 156, 102920 (2024).
\bibitem{Tx4} J. Sadeghi, M. A. S. Afshar, S. N. Gashti, M. R. Alipour, Phys. Scripta 99, 025003 (2024).
\bibitem{Tx5} J. Sadeghi, M. R. Alipour, S. N. Gashti, M. A. S. Afshar, arXiv: 2306.16117[gr-qc].



\bibitem{k1} A. Sen, Phys. Rev. Lett. 69, 1006 (1992).
\bibitem{k2} T. Houri, D. Kubiznak, C. M. Warnick and Y. Yasui, JHEP 07, 055 (2010).
\bibitem{k3} S. Q. Wu and X. Cai, J. Math. Phys. 44, 1084 (2003).
\bibitem{Ry} A. R\'{e}nyi, Acta Math. Acad. Sci. Hung. 10, 193 (1959).
\bibitem{tt} P. Cunha, V. P. and C. A. R. Herdeiro, Phys. Rev. Lett. 124, 181101 (2020).
\bibitem{T27} Y. S. Duan, M. L. Ge, Sci. Sin. 9, 1072 (1979).
\bibitem{T28} Y. S. Duan, Report No. SLAC-PUB-3301 (1984).
\bibitem{T29} Y. S. Duan, S. Li, G. H. Yang, Nucl. Phys. B 514, 705 (1998).
\bibitem{ka1} Di. Wu, Phys. Rev. D 102, 044007 (2020).

\end{thebibliography}
\end{document}